\documentclass[camera, 10pt, pageno]{jpaper}
\usepackage{cite}
\usepackage{placeins}
\usepackage{pifont}
\usepackage{color}
\usepackage{amsmath}
\usepackage{amssymb}
\usepackage{amsthm}
\usepackage{bm}
\usepackage{macros}
\usepackage{setspace}
\usepackage{xspace}
\usepackage{caption}
\usepackage{subcaption}
\usepackage{url}
\usepackage{hhline}
\usepackage{footnote}
\usepackage{refcount}
\makesavenoteenv{tabular}
\makesavenoteenv{table}

\newcommand{\nameTxt}{Focus}
\newcommand{\name}{\sf Focus}
\newcommand{\nameSec}{\large\sf Focus}

\newcommand{\panictwo}[1]{\vspace{-#1 plus 2pt minus 2pt}}
\newcommand{\ncaption}[1]{
  \panictwo{4pt}
  \renewcommand{\baselinestretch}{0.95}
  \caption{\small \bf #1}
  \panictwo{2pt}
  \renewcommand{\baselinestretch}{1}
}

\newcommand{\xref}[1]{\S\ref{#1}}

\newcommand{\kh}[1]{\textcolor{purple}{#1}}

\newcommand{\tr}[1]{#1}

\begin{document}



\title{Focus: Querying Large Video Datasets with Low Latency and Low
  Cost}

\newcommand{\cmu}{{\large$^\dag$}}
\newcommand{\msft}{{\large$^\mathsection$}}
\newcommand{\ethz}{{\large$^*$}}

\author{\rm{Kevin Hsieh\cmu\msft \quad
    Ganesh Ananthanarayanan\msft \quad
    Peter Bodik\msft \quad
    Paramvir Bahl\msft \quad
    Matthai Philipose\msft \quad
    }\\
\rm{
Phillip B. Gibbons\cmu \quad
Onur Mutlu\ethz\cmu}
 \vspace{1mm} \\
\rm{\textbf{\cmu Carnegie Mellon University \quad \msft Microsoft \quad \ethz ETH Z{\"u}rich}}
}

\maketitle
\begin{abstract}

Large volumes of videos are continuously recorded from cameras deployed for traffic control and surveillance with the goal of answering ``after the fact'' queries: {\em identify video frames with objects of certain classes (cars, bags)} from many days of recorded video. While advancements in convolutional neural networks (\cnns) have enabled answering such queries with high accuracy, they are too expensive and slow. We build {\name}, a system for low-latency and low-cost querying on large video datasets. {\name} uses  cheap ingestion techniques to index the videos by the objects occurring in them. At ingest-time, it uses compression and video-specific specialization of \cnns. \tr{{\name} handles} the lower accuracy of the cheap \cnns by judiciously leveraging expensive \cnns at query-time. To reduce query time latency, we cluster similar objects and hence avoid redundant processing. Using experiments on video streams from traffic, surveillance and news channels, we see that {\name} uses $58\times$ fewer GPU cycles than running expensive ingest processors and is $37\times$ faster than processing all the video at query time.

\end{abstract}

\section{Introduction}
\label{sec:intro}

Cameras are ubiquitous, with millions of them deployed by government and private entities at traffic intersections, enterprise offices, and retail stores. Videos from these cameras are continuously recorded \cite{genetec, avigilon}. One of the main purposes for recording the videos is answering ``after-the-fact'' queries: {\em identify video frames with objects of certain classes (like cars or bags)} over many days of recorded video. As results from these queries are used by analysts and investigators, achieving low query latencies is crucial. 



Advances in convolutional neural networks (\cnns) backed by copious training data and hardware accelerators (e.g., GPUs~\cite{k80}) have led to high accuracy in the computer vision tasks like object detection and object classification. For instance, the ResNet$152$ object classifier \cnn~\cite{DBLP:conf/cvpr/HeZRS16} won the ImageNet challenge that evaluates classification accuracy on $1,000$ classes using a public image dataset with labeled ground truths \cite{ILSVRC15}. 
For each image, these classifiers return a ranked list of $1,000$ classes in decreasing order of confidence.

Despite the accuracy of image classifier \cnns (like ResNet152), using them for video analytics queries is both expensive and slow. Using the ResNet$152$ classifier at {\em query-time} to identify video frames with cars on a month-long traffic video requires $280$ GPU hours and costs $\$250$ in the Azure cloud. The latency for running queries is also high. To achieve a query latency of one minute on $280$ GPU hours of work would require tens of thousands of GPUs classifying the frames of the video in parallel, which is many orders of magnitude more than what is typically provisioned (few tens or hundreds) by traffic jurisdictions or retail stores. 
Note that the above cost and latency values are {\em after} using motion detection techniques to exclude frames with no moving objects.

We believe that enabling {\em low-latency and low-cost querying over large video datasets} will make video analytics more useful and open up many new opportunities.

\begin{figure}[t]
  \centering
  \includegraphics[width=0.48\textwidth]{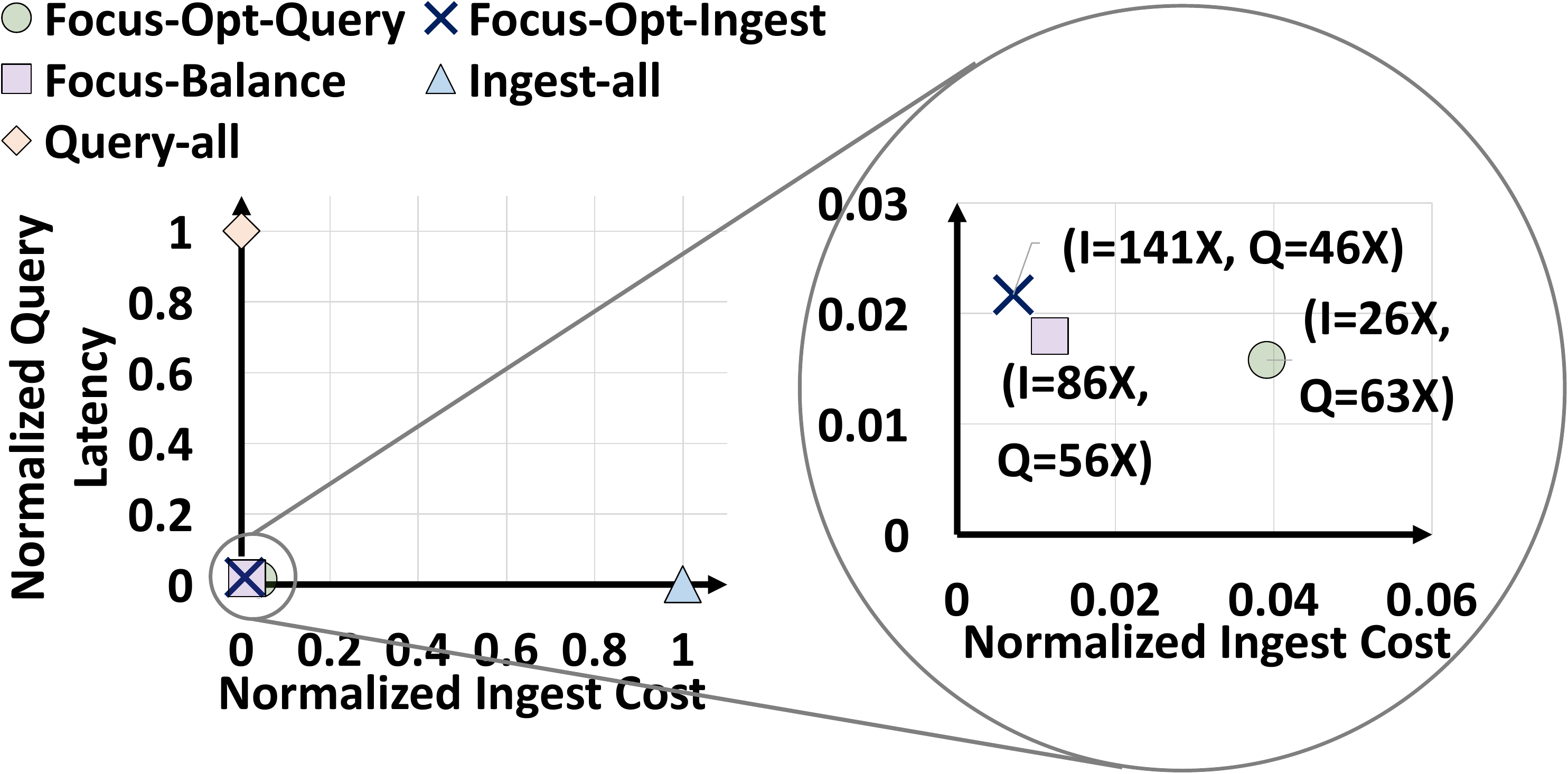}
  \caption{Effectiveness of {\name} at reducing both ingest cost and query latency, for an example traffic video.  We compare
against two baselines: \sys{``Ingest-all''} that runs ResNet152 on all video frames during ingest, and \sys{``Query-all''} that
runs ResNet152 on all the video frames at query time.
By zooming in, we see that
{\name} (the {\name-Balance} point) is simultaneously 86$\times$ cheaper
than \sys{Ingest-all} in its GPU consumption {\em and}
56$\times$ faster than \sys{Query-all} in query
latency, all the while achieving at least 95\% precision and recall.
(Also shown are two alternatives offering slightly different trade-offs.)
  }
  \label{fig:result_trade_off}
\end{figure}

A natural approach to enabling low latency querying is doing all classifications with ResNet152 at \emph{ingest-time}, i.e., on the \emph{live} videos, and store the results in an index of object classes to video frames. Any queries for specific classes (e.g., cars) will thus involve only a simple index lookup at {\em query-time}. There are, however, at least two problems with this approach. First, the cost to index all the video at ingest-time, e.g., \$250/month/stream in the above example, is prohibitively high. Second, most of this ingest-time cost is wasteful because typically only a small fraction of recorded videos get queried \cite{video-trends}. Following a theft, the police would query a few days of video from a handful of surveillance cameras, but not all the videos.

We present {\name}, a system to support low-latency low-cost querying on large video datasets.
To address the above drawbacks, {\name} has the following goals:
$(1)$ low cost indexing of video at ingest-time,
$(2)$ providing high accuracy and low latency for queries, and
$(3)$ allowing trade offs between the cost at ingest-time against the latency at query-time.
As input, the user specifies the \emph{ground-truth CNN} (or ``GT-CNN'', e.g., the ResNet152 classifier) and the desired accuracy of results that {\name} needs to achieve relative to the GT-CNN.

{\name} uses four key techniques -- cheap \cnns for ingest, using top-{\sf K} results from the ingest-time \cnn, clustering similar objects, and judicious
selection of system and model parameters.





First, to make video ingestion cheap, {\name} uses \emph{compressed} and \emph{specialized} versions of \cnns, to create an ingest-time index of object classes to frames.
CNN compression \tr{(e.g.,~\cite{Simonyan15})} creates new CNNs with fewer convolutional layers and smaller input images.
Specialization\tr{~\cite{mcdnn, DBLP:journals/corr/ShenHPK17}} trains those CNNs on a smaller set of
object classes specific to each video stream \tr{so that those
cheaper CNNs can classify these video-specific objects more accurately}.
Together, these techniques result in highly efficient CNNs for video indexing.



Second, the cheap ingest \cnns, however, are also less accurate than the expensive GT-\cnn (like ResNet$152$), measured in terms of {\em recall} and {\em precision}. Recall is the fraction of frames in the video that contained objects of the queried class that were {\em actually} returned in the query's results. Precision, on the other hand, is the fraction of frames in the query's results that contained objects of the queried class.
To increase recall, {\name} relies on an empirical observation: while the top-most (i.e., most confident) classification results of the cheap and expensive \cnns may not always match, the top-most result of the expensive \cnn falls within the {\em top-{\sf K}} results of the cheap \cnn. Therefore, at ingest-time, {\name} indexes each object with the ``top-{\sf K}'' results of the cheap CNN (instead of just the top-most).
To increase precision, at query-time, we first filter the objects from the top-{\sf K} index and then classify the filtered objects with the expensive GT-\cnn.

Third, to reduce the query-time latency of using the expensive GT-\cnn, {\name} relies on the significant similarity between objects in videos.
For example, a car moving across an intersection will look very similar in consecutive frames.
{\name} leverages this similarity by clustering the objects at ingest-time,
classifying {\em only} the cluster centroids with the expensive GT-\cnn at query-time, and assigning the same class to all objects in the cluster, thus considerably reducing query latency.

In a nutshell, {\name}'s ingest-time and query-time operations are as follows. At ingest-time, it classifies the detected objects using a cheap \cnn, clusters similar objects, and indexes each cluster centroid using the top-K classification results.
At query-time, when the user queries for class X, {\name} looks up the ingest index for centroids that match class X and classifies them using the GT-CNN. For centroids that were classified as class X, it returns all objects from the corresponding clusters to the user.



Finally, {\name} smartly chooses the ingest-time \cnn and its parameters to meet user-specified targets on precision and recall.
Among the choices that meet the accuracy targets, it allows the user to trade off between the ingest cost and query latency.
For example, using a cheaper ingest \cnn reduces the
ingest cost but increases the query latency \tr{as {\name} needs to
  use a larger {\sf K} for the top-{\sf K} index to retain the
  accuracy targets}.
{\name} identifies the ``sweet spot'' in parameters that sharply improve one of ingest cost or query latency for a small worsening of the other.

We built {\name} and evaluated it on thirteen 12-hour videos from three domains --
traffic cameras, surveillance cameras, and news channels. We compare
against two baselines: \sys{``Ingest-all''} that runs GT-CNN on all video frames during ingest, and \sys{``Query-all''} that
runs GT-CNN on all the video frames at query time. We use ResNet152 as GT-CNN and augment both
baselines with motion detection to remove frames with no
objects, \tr{which is one of the core techniques in a recent prior work, NoScope~\cite{DBLP:journals/pvldb/KangEABZ17}.}
Figure~\ref{fig:result_trade_off} shows a representative result,
for a traffic video from a commercial intersection. 
On average, {\name} is $58\times$ (up to 98$\times$) cheaper
than \sys{Ingest-all} and $37\times$ (up to 57$\times$) faster than
\sys{Query-all}. This leads to the cost of ingestion coming down from \$250/month/stream to \$4/month/stream, and the latency to query a 24 hour video dropping from $1$ hour to under $2$ minutes. 
See \xref{sec:evaluation} for the full details.



We make the following contributions.
\begin{enumerate}
\item We formulate the problem of querying video datasets by showing the trade-offs between query latency, ingest cost, and accuracy (precision and recall) of results.
\item \tr{We propose techniques to ingest videos with low cost by leveraging compressed and video-specific specialization of \cnns, while retaining high accuracy targets by creating approximate (top-{\sf K}) indexes.}
\item We identify and leverage similarity between objects in a video to cluster them using \cnn features and significantly speeding up queries.
\item \tr{We propose and build a new end-to-end system to support low-latency, low-cost querying on large video datasets. We show that our system offers new trade-off options between ingestion cost and query latency, as it is significantly cheaper than analyzing all videos frames at ingest time and significantly faster than analyzing queried video frames at query time. }  
\end{enumerate}

\section{Background and Motivation}
\label{sec:background}

We first provide a brief overview of convolutional Neural Networks (\cnn), the state-of-the-art approach to detecting and classifying objects in images (\xref{subsec:cnn}).
We then discuss new observations we made about real-world videos, which motivate the design of our techniques (\xref{subsec:char}).

\subsection{Convolutional Neural Networks}
\label{subsec:cnn}

A Convolution Neural Network
(\cnn)~\cite{DBLP:journals/neco/LeCunBDHHHJ89} is a specific class of neural networks that works by extracting the visual features in images.
During image classification, or ``inference'', a \cnn takes an input image and outputs the probability of each \emph{class} (e.g., dog, flower, or car).
\cnns are the state-of-the-art method used for many computer
vision tasks, such as image classification
(e.g.,~\cite{DBLP:conf/nips/KrizhevskySH12,
  DBLP:conf/cvpr/SzegedyLJSRAEVR15, DBLP:conf/cvpr/HeZRS16}) and face
recognition (e.g.,~\cite{DBLP:journals/tnn/LawrenceGTB97,
  DBLP:conf/cvpr/SchroffKP15}).

\begin{figure}[h]
  \centering
  \includegraphics[width=0.48\textwidth]{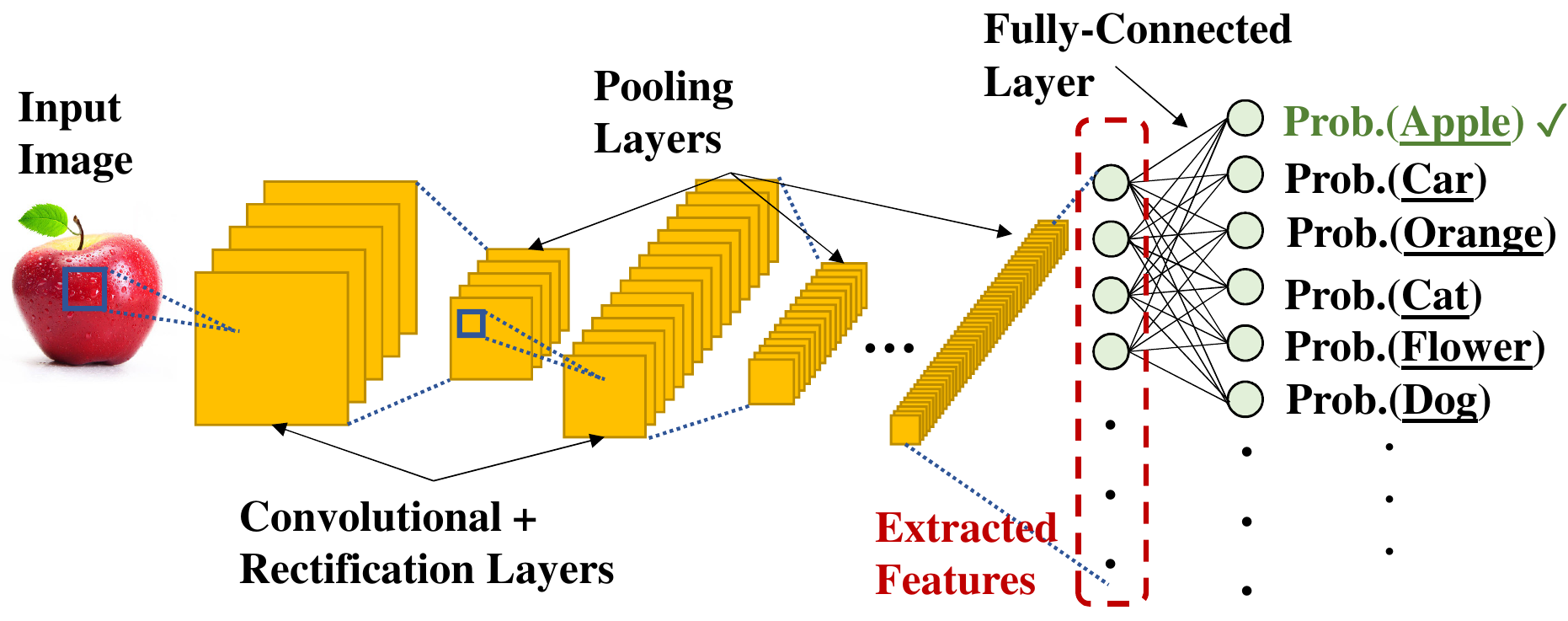}
  \caption{Architecture of an image classification \cnn.}
  \label{fig:cnn}
\end{figure}

Figure~\ref{fig:cnn} illustrates the architecture of an
image classification \cnn. Broadly, almost all \cnns consist of three key
types of network layers: (1) {\em convolutional and rectification
layers}, which detect visual features from input pixels, (2) {\em pooling layers}, which down-sample the input by
merging neighboring pixel values, and (3) {\em fully-connected layers},
which provide the reasoning to classify the input
object based on the outputs from previous layers.
The outputs of an image classification \cnn are the the probabilities of all object classes, and the class with the highest probability is the predicted class for the input image.

The output of the penultimate (i.e., previous-to-last) layer can be considered as ``representative features'' of
the input image~\cite{DBLP:conf/nips/KrizhevskySH12}. The features are a real-valued vector, with lengths between $512$ and $4096$ in state-of-the-art classifier \cnns
(e.g.,~\cite{DBLP:conf/nips/KrizhevskySH12, Simonyan15, DBLP:conf/cvpr/SzegedyLJSRAEVR15, DBLP:conf/cvpr/HeZRS16}).
It has been shown that images with
similar feature vectors (i.e., small Euclidean distances) are visually
similar~\cite{DBLP:conf/nips/KrizhevskySH12,
  DBLP:conf/eccv/BabenkoSCL14, DBLP:conf/cvpr/RazavianASC14, DBLP:conf/iccv/BabenkoL15}.


The high accuracy of \cnns comes at a cost: {\em inferring} (or classifying) using state-of-the-art \cnns
to classify objects in images requires significant
computational resources.
This is because the higher accuracy of \cnns comes from using \emph{deeper} architectures (i.e., more layers)
to obtain better visual features.
For instance, ResNet152~\cite{DBLP:conf/cvpr/HeZRS16}, the
winner of the ImageNet competition~\cite{ILSVRC15} in $2015$, has been trained to classify across 1000 classes from the ImageNet dataset using $152$ layers, but can only process $77$ images/second even with a high-end GPU (NVIDIA K80~\cite{k80}).
This makes querying on large video datasets using these \cnns to be
\emph{slow} and \emph{costly}.



There are at least two recent techniques designed to reduce the cost of \cnns.
First, \emph{compression} is a set of techniques aiming to reduce the cost of \cnn inference (classification) at the expense of reduced accuracy.
Such techniques include removing some expensive convolutional layers~\cite{Simonyan15}, matrix pruning~\cite{pruning1,pruning2}, and others~\cite{lowrank,fitnets} and can dramatically reduce the classification cost of a \cnn.
For example, ResNet18, which is a ResNet152 variant with only 18 layers is $8\times$ cheaper.
Second, a more recent technique is \cnn \emph{specialization}~\cite{mcdnn}, where the \cnns are trained on a subset of a dataset specific to a particular context, also making them much cheaper.
Using the combination of cheap and expensive \cnns is a key facet of our solution, described in \xref{sec:techniques}.





\subsection{Characterizing Real-world Videos}
\label{subsec:char}

We aim to support queries of the form, {\em find all frames in the video that contain objects of class X}. We identify some key characteristics of real-world videos towards supporting these queries: $(1)$ large portions of videos can be excluded (\xref{subsubsec:filtering}), $(2)$ only a limited set of object classes occur in each video (\xref{subsubsec:limited}), and $(3)$ objects of the same class have similar feature vectors (\xref{subsubsec:features}). The design of {\name} is based on these characteristics.

We have analyzed 12 hours of video from six video streams each. The six video stream span across traffic cameras, 
surveillance cameras, 
and news channels. 
(\xref{sec:methdology} provides the details.) We detect the objects in each frame of these videos (using background subtraction~\cite{DBLP:conf/avbs/KaewTraKulPongB15}), and classify each object with the ResNet152 \cnn \cite{DBLP:conf/cvpr/HeZRS16} among the supported $1,000$ object classes. In this paper, we use results from the costly ResNet152 \cnn as ground truth.

\subsubsection{Excluding large portions of videos}
\label{subsubsec:filtering}

We find considerable potential to avoid processing large portions of videos at query-time. 
Significant portions of video streams either have {\em no objects} at all (as in a garage camera at night) or the objects are {\em stationary} (like parked cars). We find that in our video sets, one-third to one-half of the frames fall in these categories. Therefore, queries to {\em any} object class would benefit from pre-processing filters applied to exclude these portions of the videos. 
Even among the frames that do contain objects, not all of them are relevant to a query because each query only looks for a {\em specific class} of objects. 
In our video sets, an object class occurs in only $0.01$\% of the frames on average, and even the most frequent object classes occur in no more than $16\%-43\%$ of the frames in the different videos. This is because while there are usually some dominant classes (e.g., cars in a traffic camera, people in a news channel), most other classes are rare.
Since queries are for specific object classes, there is considerable potential in {\em indexing} frames by the classes of objects.



\begin{figure}[t!]
  \centering
  \includegraphics[width=0.43\textwidth]{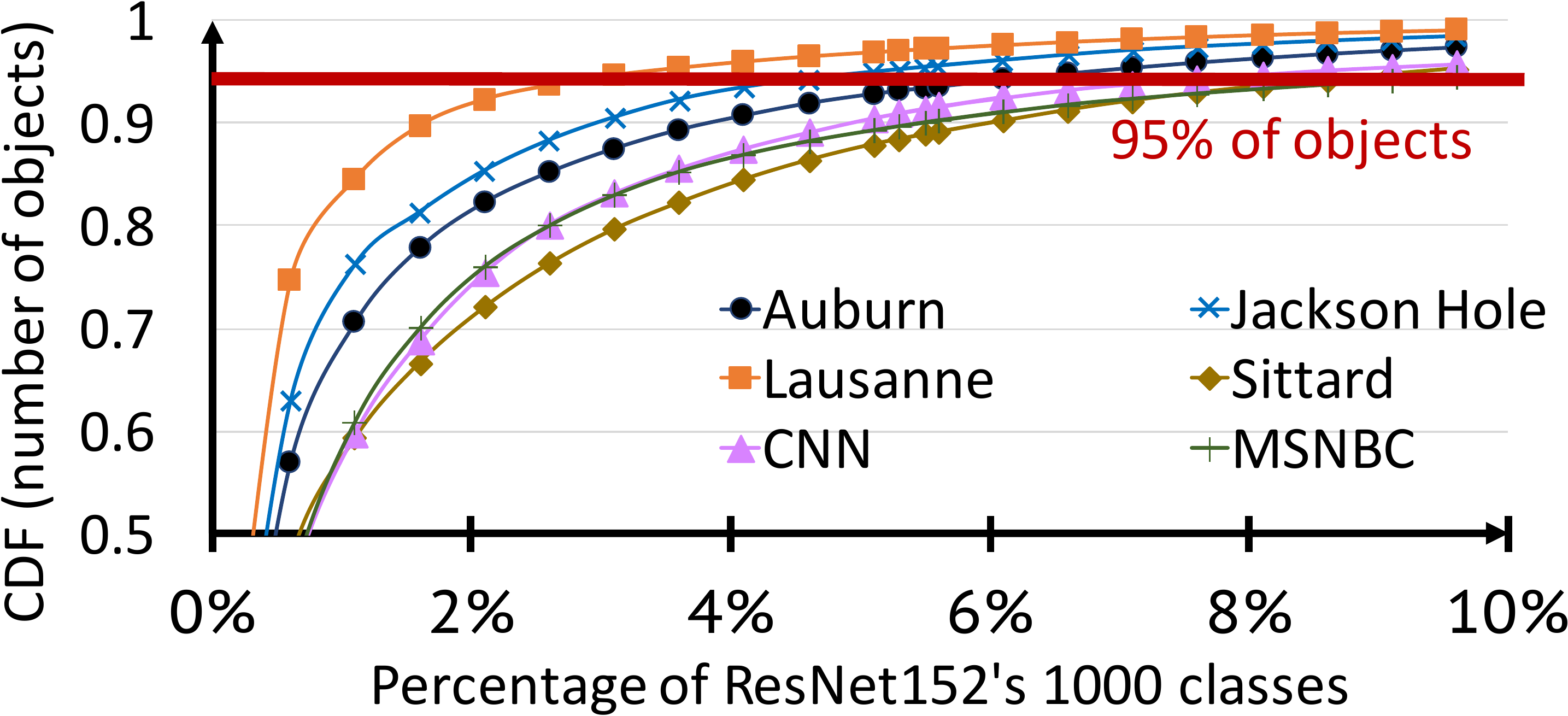}
  \caption{CDF of frequency of object classes. The x-axis is the fraction of classes out of the $1000$ recognized by ResNet152 (truncated to $10\%$).}
  \label{fig:class_num_cdf}
\end{figure}

\begin{figure*}[t!]
\centering
\includegraphics[width=0.9\textwidth]{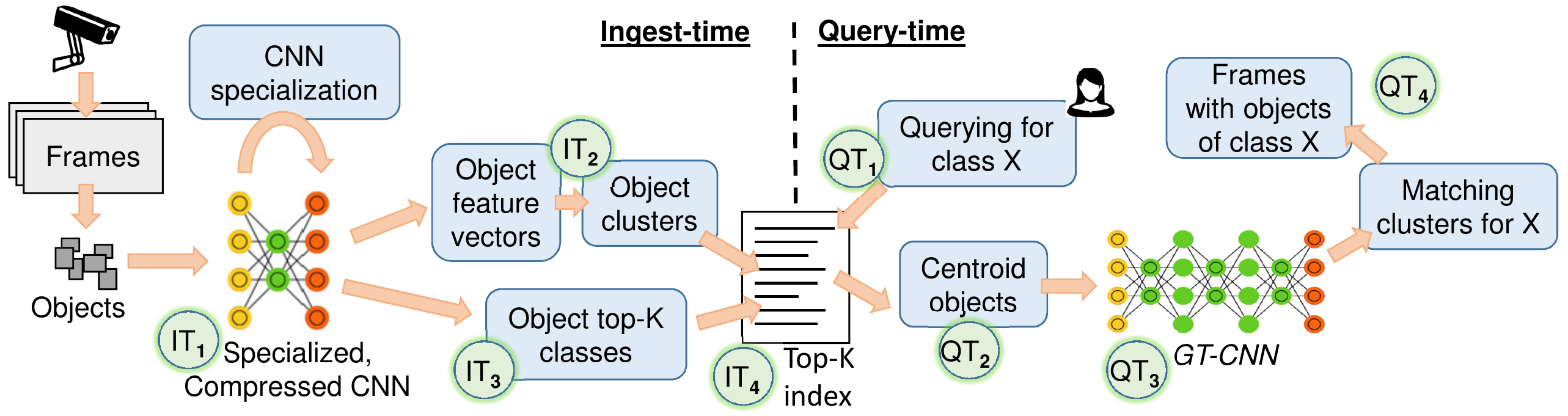}
\vspace{0.1in}
\ncaption{Overview of {\name}.}
\vspace{0.1in}
\label{fig:arch}
\end{figure*}

\subsubsection{Limited set of object classes in each video}
\label{subsubsec:limited}

We next focus on the classes of objects that occur in each of the videos and the disparity in frequency among them.

Most video streams have a limited set of objects because each video has its own context (e.g., traffic cameras can have automobiles, pedestrians or bikes but not airplanes). It is rare that a video stream contains objects of {\em all} the classes recognized by state-of-the-art classifier \cnns. Figure~\ref{fig:class_num_cdf} shows the cumulative distribution function (CDF) of the frequency of object classes in our videos (as classified by ResNet152). We make two observations.

First, objects of only $22\%-33\%$ (not graphed) of the $1,000$ object
classes occur in the less busy videos (\texttt{Auburn},
\texttt{Jackson Hole}, \texttt{Lausanne}, and \texttt{Sittard}). Even
in the busier videos (\texttt{CNN}, and \texttt{MSNBC}), objects of
only $50\%-69\%$ of the classes appear. Also, there is little overlap
between the classes of objects among the different videos. On average,
the Jaccard indexes~\cite{jaccard} (i.e., intersection over union)
between the videos based on their object classes is only
$0.46$. Second, even among the object classes that do occur, a small
fraction of classes disproportionately
dominate. Figure~\ref{fig:class_num_cdf} shows that $3\%-10\%$ of the
most frequent object classes cover $\geq 95\%$ of the objects in each
video stream.  This suggests that for each video stream, we can
\textit{automatically} (i) determine its most frequently occurring
classes and (ii) train efficient \cnns \emph{specialized} for classifying
these classes (\xref{subsec:cnn}).



\subsubsection{Feature vectors for finding duplicate objects}
\label{subsubsec:features}

Objects moving in the video often stay in the frame for several seconds; for example, a pedestrian might take a minute to cross a street.
Instead of classifying {\em each instance} of the same object across the frames, we would like to \emph{inexpensively} find duplicate objects and only classify one of them using a \cnn (and apply the same label to all duplicates). Thus, given $n$ duplicate objects, this requires only one \cnn classification operation instead of $n$.

Comparing pixel values across frames is an obvious choice to identify duplicate objects, however, they turn out to be highly sensitive to even small changes in the camera's real-time view of an object. Instead,
feature vectors extracted from the \cnns are much more robust since they are specifically trained to extract visual features for classification.
We verify the robustness of feature vectors using the following analysis. 
In each video, for each object $i$, we find its \emph{nearest} neighbor $j$ using feature vectors from the cheap ResNet18 \cnn and compute the fraction of object pairs that belong to the same class.
This fraction is over 99\% in each of our videos, which shows using feature vectors from cheap \cnns can potentially help identify duplicate objects.

\section{Overview of {\nameSec}}
\label{sec:architecture}


The goal of {\name} is to {\em index live video streams} by the object classes occurring in them and enable answering ``after-the-fact'' queries later on the stored videos of the form \emph{find all frames that contain objects of class X}. Optionally, the query can be restricted to a subset of cameras and a time range. Such a query formulation is the basis for many widespread applications and could be used either on its own (such as for detecting all cars or bicycles in the video) or used as a basis for further processing (e.g., finding all collisions between cars and bicycles).

{\name} is designed to work with a wide variety of current and future
\cnn{}s.  At system configuration time, the user (system
administrator) provides a \emph{ground-truth CNN} (GT-\cnn), which
serves as the accuracy baseline for {\name}, but is far too costly to
run on every video frame.  Through a sequence of techniques, {\name}
provides nearly-comparable accuracy but at
greatly reduced cost.  By default, and throughout this paper, we use
the ResNet152 image classifier as the GT-\cnn.

Because the acceptable target accuracy is application-dependent, {\name} permits the user to specify the target, while providing reasonable defaults.
Accuracy is specified in terms of \emph{precision}, i.e., fraction of frames output by the query that actually contain an object of class X according to GT-\cnn, and \emph{recall}, i.e., fraction of frames that contain objects of class X according to GT-\cnn that were actually returned by the query.
The lower the target, the greater the cost-savings provided by {\name}.
Even for high targets such as 95\%--99\%,
{\name} is able to achieve order-of-magnitude or more cost savings.
Figure~\ref{fig:arch} presents the design of {\name}.
\begin{itemize}
\item
At \emph{ingest-time} (left part of Figure~\ref{fig:arch}), {\name} classifies objects in the incoming video frames and extracts their feature vectors.
To make this step cheap, it uses a highly compressed and specialized version of the GT-\cnn model (IT$_\text{1}$ in Figure~\ref{fig:arch}). 
{\name} then clusters objects based on their feature vectors (IT$_\text{2}$) and assign to each cluster the \emph{top {\sf K}} most likely classes these objects belong to (based on classification confidence of the ingest \cnn); (IT$_\text{3}$). 
It creates a \emph{top-{\sf K} index}, which maps each class to the set of object clusters (IT$_\text{4}$). The top-{\sf K} index is the output of {\name}' ingest-time processing of videos.
\item
At {\em query-time} (right part of Figure~\ref{fig:arch}), when the user queries for a certain class $X$ (QT$_\text{1}$), {\name} retrieves the matching clusters from the top-{\sf K} index (QT$_\text{2}$), runs the \emph{centroids} of the clusters through GT-\cnn (QT$_\text{3}$), and returns all frames from the clusters whose centroids were classified by GT-\cnn as class $X$ (QT$_\text{4}$). 
\end{itemize}


\noindent{The top-{\sf K} ingest index is a mapping between the object class to the clusters. Specifically,

\texttt{object class $\rightarrow$ $\langle$cluster ID$\rangle$}

\texttt{cluster ID $\rightarrow$ [centroid object, $\langle \texttt{objects}\rangle$ in cluster, $\langle \texttt{frame IDs}\rangle$ of objects]}

We next explain how {\name}' key techniques keep ingest cost and query latency low while also meeting the user-specified accuracy targets. 

\noindent{\bf 1) Cheap Ingest-time \cnn:} {\name} makes indexing at ingest-time cheap by compressing and specializing the GT-\cnn model for {\em each} video stream. $(i)$ {\em Compression} of \cnn models\tr{~\cite{Simonyan15, pruning1,pruning2,lowrank,fitnets}}
uses fewer convolutional layers and other approximation techniques (\xref{subsec:cnn}). $(ii)$ {\em Specialization} of \cnns\tr{~\cite{mcdnn, DBLP:journals/corr/ShenHPK17}} uses the observation that a specific video stream contains only a small number of object classes and their appearance is more constrained than in a generic video (\xref{subsubsec:limited}). Both techniques are done automatically and together result in ingest-time \cnn models that are up to $98\times$ cheaper than GT-CNN.

\noindent{\bf 2) Top-{\sf K} ingest index:} The cheap ingest-time \cnns are less accurate, i.e., their top-most results do not often match the top-most classifications of GT-\cnn. 
Therefore, to keep the recall high, 
{\name} associates each object with the \emph{top-{\sf K}} classification results of the cheap \cnn, instead of just its top-most result.
Increasing the {\sf K} increases recall because the top-most result of GT-\cnn often falls within the ingest-time \cnn's top-{\sf K} results. 
At query-time, {\name} uses the GT-\cnn to remove objects in this larger
set that do not match the class, to regain precision lost by including
all the top-{\sf K}.

\noindent{\bf 3) Clustering similar objects:} A high value of {\sf K} at ingest-time increases the work to do at query time, thereby increasing query latency. To reduce this overhead, {\name} clusters similar objects at ingest-time using feature vectors from the ingest-time \cnn. In each cluster, at query-time, we run only the cluster centroid through GT-\cnn and apply the classified result from the GT-\cnn to all objects in the cluster. Thus, if the objects are not tightly clustered, clustering can reduce precision and recall.




\noindent{\bf 4) Trading off ingest vs.~query costs:} {\name} automatically chooses the cheap \cnn, its {\sf K}, and specialization and clustering parameters to achieve the desired precision and recall targets. These choices also help {\name} trade off between the work done at ingest-time and query-time.
For instance, to save ingest work, {\name} can select a cheaper ingest-time \cnn,
and then counteract the resultant loss in accuracy by running the
expensive GT-\cnn on more objects at query time.
{\name} chooses its parameters so as to offer a sharp improvement in one of the two costs for a small degradation in the other cost.
Because the desired trade-off point is application-dependent,
{\name} provides users with a choice of three options:
ingest-optimized, query-optimized, and balanced (the default).


Note that while our explanation is anchored on image classification \cnns,  the architecture of {\name} is generally applicable to all existing \cnns (e.g., face recognition). Techniques that we use for \cnn compression~\cite{lowrank,fitnets} and specialization~\cite{mcdnn}, and feature extraction from the \cnns are all broadly applicable to all \cnns.

\section{Video Ingest \& Querying Techniques}
\label{sec:techniques}

In this section, we describe the main techniques used in {\name}: using cheap \cnn models at ingest-time (\xref{subsec:cheap_ingest}), identifying similar objects and frames to save on redundant \cnn processing (\xref{subsec:redundancy}), and specializing the \cnns to the specific videos that are being analyzed (\xref{subsec:specialization}). 
\xref{subsec:tuning} describes setting parameters in {\name}.



\subsection{Cheap Ingestion}
\label{subsec:cheap_ingest}


{\name} indexes the live videos at {\em ingest-time} to reduce the {\em query-time} latency. We perform object detection on each frame, typically an inexpensive operation, and then will classify the extracted objects using {\em ingest-time} \cnns that are far cheaper than the ground-truth GT-\cnn. We use these classifications to index objects by class.

\noindent{\bf Cheap ingest-time \cnn:}
As noted earlier, the user provides {\name} with a GT-CNN.
Optionally, the user can also provide other classifier architectures to be
used in {\name}' search for cheap \cnns, such as
AlexNet \cite{DBLP:conf/nips/KrizhevskySH12} and VGG \cite{Simonyan15} (which vary in their resource costs and accuracies).
Starting from these user-provided CNNs, {\name} applies various levels of compression, such as removing convolutional layers and reducing the input image resolution (\xref{subsec:cnn}).
This results in a large set of \cnn options for ingestion, \{CheapCNN$_1, \dots,$ CheapCNN$_n$\}, with a wide range of costs and accuracies.

\noindent{\bf Top-{\sf K} Ingest Index:} To keep recall high, {\name} indexes each object using the {\em top {\sf K}} object classes from CheapCNN$_i$'s output, instead of using just the top-most class. Recall from \xref{subsec:cnn} that the output of the \cnn is a list of object classes in descending order of confidence. We empirically observe that the top-most output of the expensive GT-\cnn is often contained within the top-{\sf K} classes output by the cheap \cnn (for a small value of {\sf K} relative to the $1,000$ classes recognized by the \cnns).

Figure~\ref{fig:topK} plots the effect of {\sf K} on recall on one of our video streams, \video{lausanne} (see \xref{sec:methdology}).
The three models in the figure are ResNet18~\cite{DBLP:conf/cvpr/HeZRS16}, and ResNet18 with 3 and 5 layers removed; additionally, the input images were rescaled to 224, 112, and 56 pixels, respectively.
All models were retrained on their original training data (ImageNet~\cite{ILSVRC15}). We make two observations.


First, we observe steady increase in recall with increasing {\sf K}, for all three CheapCNNs.  As the figure shows, CheapCNN$_1$, CheapCNN$_2$, and CheapCNN$_3$ reach 90\% recall when {\sf K} $\geq$ 60, {\sf K} $\geq$ 100, and {\sf K} $\geq$ 200, respectively. Note that all these models recognize 1000 classes, so even {\sf K} $=$ 200 represents only 20\% of the possible classes. 
Second, there is a \emph{trade-off} between different models -- the cheaper they are, the lower their recall with the same {\sf K}.
Overall, we conclude that by selecting the appropriate {\sf K}, {\name} can achieve the target recall.


\begin{figure}[t]
  \centering
  \includegraphics[width=0.48\textwidth]{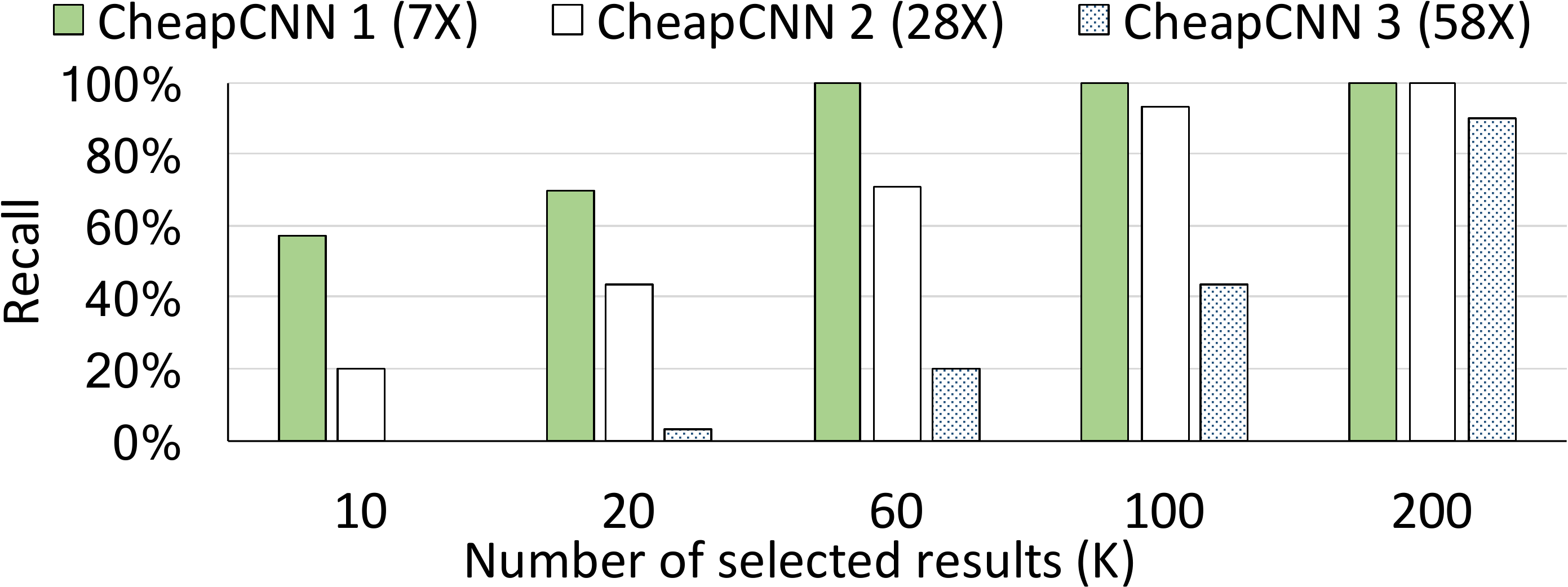}
  \caption{Effect of {\sf K} on recall for three cheap \cnns. The number within the parenthesis indicates how much cheaper the model is compared to our GT-\cnn, ResNet152.}
  \label{fig:topK}
\end{figure}

{\name} creates the {\em top-{\sf K} index} of an object's top-{\sf K} classes output by CheapCNN$_i$ at ingest-time. 
While filtering for objects of the queried class $X$ using the top-{\sf K} index (with the appropriate {\sf K}) will have a high recall, it will have very low precision. Since we associate each object with {\sf K} classes (while it has only one true class), the average precision is only $1/{\sf K}$.
Thus, at query time, to keep the precision high, {\name} determines the {\em actual} class of objects from the top-{\sf K} index using the expensive GT-\cnn and only return objects that match the queried class.


The selection of the cheap ingest-time \cnn model (CheapCNN$_i$) and the {\sf K} value (for the top-{\sf K} results) has a significant influence on the recall of the outputs produced. Lower values of {\sf K} reduce recall, i.e., {\name} will miss returning frames that contain the queried objects. At the same time, higher values of {\sf K} increase the number of objects to classify with GT-\cnn at query time to keep precision high, and hence adds to the latency. 
We defer to \xref{subsec:tuning} on how {\name} sets these parameters as they have to be jointly set with other parameters in \xref{subsec:redundancy} and \xref{subsec:specialization}.

\subsection{Redundancy Elimination}
\label{subsec:redundancy}

At query time, {\name} retrieves the objects likely matching the user-specified class from the top-{\sf K} index and infers their actual class using the GT-\cnn.
This would ensure precision of 100\%, but could cause significant latency at query time.
Even if this inference is parallelized across many GPUs, it would still incur a large cost.
{\name} uses the following observation to reduce this cost: if two objects are visually similar, their feature vectors would be closely aligned and they would likely be classified as the same class (e.g., ``cars'') by the GT-\cnn model (\xref{subsubsec:features}).

{\name} {\em clusters} objects that are similar, invokes the expensive GT-\cnn only on the cluster centroids, and assigns the centroid's label to all objects in each cluster.
Doing so dramatically reduces the work done by the GT-\cnn classifier at query time. {\name} uses the feature vector output by the previous-to-last layer of the cheap ingest \cnn (see \xref{subsec:cnn}) for clustering. Note that {\name} clusters the {\em objects} in the frames and not the frames as a whole. The key questions regarding clustering are \emph{how} do we cluster (algorithm) and \emph{when} do we cluster (system). We discuss both these key questions below.


\noindent{\bf Clustering Heuristic:} We require two properties in our clustering technique. First, given the high volume of video data, it should be a single-pass algorithm to keep the overhead low, as the complexities of most clustering algorithms are \emph{quadratic}. Second, it should make no assumption on the number of clusters and adapt to outliers in data points on the fly. Given these requirements, we use the following simple approach for {\em incremental} clustering, which has been well-studied in the literature \cite{clustering1, clustering2}.


We put the first object into the first cluster $c_1$.
To cluster a new object $i$ with a feature vector $f_i$,
we assign it to the closest cluster $c_j$ if $c_j$ is at most distance $T$ away from $f_i$.
However, if none of the clusters are within a distance $T$, we create a new cluster with centroid at $f_i$, where $T$ is a distance threshold. We measure distance as the $L_2$ norm \cite{l2norm} between cluster centroid and object feature vector.
We keep the number of clusters at a constant $M$ by removing the smallest ones and storing their data in the top-{\sf K} index.
Using this algorithm, we can keep growing the popular clusters (such as similar cars), while keeping the complexity as $O(Mn)$, which is linear to $n$, the total number of objects.

Clustering can reduce both precision and recall depending on parameter $T$.
If the centroid is classified by GT-\cnn as the queried class X but the cluster contains another object of a different class, it reduces precision.
If the centroid is classified as a class different than X but the cluster has an object of class X, it reduces recall.
We discuss setting $T$ in \xref{subsec:tuning}.

\noindent{\bf Clustering at Ingest vs. Query Time:} {\name} clusters the objects at ingest-time rather than at query-time.
Clustering at query-time would involve \emph{storing} all feature vectors, \emph{loading} them for objects filtered from the ingest index and then clustering them.
Instead, clustering {\em at ingest time} creates clusters right when the feature vectors are created and only stores the cluster centroids in the top-{\sf K} index.
This makes the query-time latency much lower and also reduces the size of the top-{\sf K} index.
We observe that the ordering of indexing and clustering operations is mostly \emph{commutative} in practice and has little impact on result accuracy (we do not present these results due to space constraints).
We therefore use ingest-time clustering due to its latency and storage benefits.


\noindent{\bf Pixel Differencing of Objects:} While clustering primarily reduces work done at query-time (number of objects to be classified by the GT-\cnn), {\name} also employs {\em pixel differencing} among objects in adjacent incoming frames to reduce ingest cost. Specifically, if two objects have very similar pixel values, it only runs the cheap \cnn on one of them and assign them both to the same cluster in our top-{\sf K} index.

%

\subsection{Video-specific Specialization of \cnns}
\label{subsec:specialization}

Recall from \xref{subsec:cheap_ingest} that {\name} uses a cheap ingest-time \cnn, CheapCNN$_i$ to index object classes. {\name} further reduces its cost by {\em specializing} the ingest-time \cnn model to each video stream.
Model specialization benefits from two properties of objects in each video stream.
First, while object classification models are trained to differentiate between thousands of object classes, many video streams contain only a small number of classes (\xref{subsubsec:limited}).
Second, objects in a specific stream are often visually more constrained than objects in general (say, compared to the ImageNet \cite{ILSVRC15} dataset). The cars and buses that occur in a specific traffic camera have much less variability, e.g., they have very similar angle, distortion and size, than a generic set of vehicles. 


Instead of trying to differentiate among thousands of object classes, differentiating among just (say) fifty classes {\em and} in a specific camera's video is a much simpler task, requiring simpler image features and smaller image resolutions. As a result, the specialized models are \emph{smaller} and \emph{more accurate}~\cite{mcdnn}. For example, by retraining a stream-specific CheapCNN$_i$, we can achieve similar accuracy on video streams, while removing $1/3$ of the convolutional layers and making the input image 4$\times$ smaller in resolution. This leads to the specialized CheapCNN$_i$ being 10$\times$ cheaper than even the generic CheapCNN$_i$.



Since the specialized \cnn classifies across fewer classes, they are more accurate, which allows {\name} to select a much smaller {\sf K} (for the top-{\sf K} ingest index) to meet the desired recall. We find that specialized models can use {\sf K} $=$ 2 or 4, much smaller than the typical {\sf K} $=$ 60 \textasciitilde~200 for the generic cheap \cnns (Figure~\ref{fig:topK}).
Smaller {\sf K} directly translates to fewer objects that have to be classified by GT-\cnn at query time, thus reducing latency.


\noindent{\bf Model Retraining:} On each video stream {\name} periodically obtains a small sample of video frames and classifies their objects using GT-\cnn to estimate the ground truth of distribution of object classes for the video (similar to Figure~\ref{fig:class_num_cdf}). From this distribution, {\name} selects the most frequently occurring $L_s$ object classes to retrain new specialized models. 
As we saw in \xref{subsubsec:limited}, there is usually a ``power law'' in the distribution of classes -- just a handful of classes account for a dominant majority of the objects -- thus, low values of $L_s$ usually suffice.\footnote{
Specialized \cnns can be retrained quickly on a small dataset. Retraining is relatively infrequent and done once every few days.}

Specialization is also based off a family of \cnn architectures (such as ResNet \cite{DBLP:conf/cvpr/HeZRS16}, AlexNet \cite{DBLP:conf/nips/KrizhevskySH12}, and VGG \cite{Simonyan15}) with different number of convolution layers, 
similar to \xref{subsec:cheap_ingest}. 
Specialization adds to the set of options available for ingest \cnns (\{CheapCNN$_1, ... ,$ CheapCNN$_n$\} in \xref{subsec:cheap_ingest}), and {\name} picks the best model (CheapCNN$_i$) and the corresponding {\sf K} for the index.


\noindent{\bf ``OTHER'' class:} While {\name} specializes the \cnn towards the most frequently occurring $L_s$ classes, we also want to support querying of the {\em less} frequent classes. For this purpose, {\name} includes an additional class called ``OTHER'' in the specialized model.\footnote{Since there will be considerably fewer objects in the video belonging to the OTHER class, we proportionally re-weight the training data to contain equal number of objects of all the classes.} Being classified as OTHER simply means not being one of the $L_s$ classes. At query time, if the queried class is among the OTHER classes of the ingest \cnn's index, {\name} extracts all the clusters that match the OTHER class and classifies their centroids through the GT-\cnn model. 

The parameter $L_s$ (for each stream) exposes the following trade-off. Using a small $L_s$ allows us to train a simpler model with cheaper ingest cost and lower query-time latency {\em for the popular classes}, however, it also leads to a larger fraction of objects falling in the OTHER class; querying for them will be expensive because all those objects will have to be classified by the GT-\cnn. Using a larger value of $L_s$, on the other hand, leads to a more expensive ingest and query-time models, but cheaper querying for the OTHER classes. We select $L_s$ next in \xref{subsec:tuning}.

\subsection{Balancing Accuracy, Latency, and Cost}
\label{subsec:tuning}

{\name}' goals of high accuracy, low ingest cost and low query latency are impacted by the parameters in {\name}' techniques --
{\sf K}, the number of top results from the ingest-time \cnn to index an object;
$L_s$, the number of popular object classes we use to create a specialized model;
CheapCNN$_i$, the specialized ingest-time cheap \cnn;
and $T$, the distance threshold for clustering objects.

The effect of these four parameters is intertwined.
All the four parameters impact ingest cost, query latency, and recall, but only $T$ impacts precision.
This is because we apply the cluster centroid's classification by GT-\cnn to all the objects in its cluster. Thus, if the clustering is not tight (i.e., high value of $T$), we lose precision.

\noindent{\bf Parameter Selection:} {\name} selects parameter values {\em per video stream}. It samples a representative fraction of frames of the video stream and classifies them using GT-\cnn for the ground truth. For each combination of parameter values, {\name} computes the expected precision and recall (using the ground truths generated by GT-\cnn) that would be achieved for each of the object classes.
To navigate the combinatorial space of options for these parameters, we adopt a two-step approach. In the first step, {\name} chooses CheapCNN$_i$, $L_s$ and $K$ using only the recall target.
In the next step, {\name} iterates through the values of $T$, the clustering distance threshold, and only select values that meet the precision target.

\noindent{\bf Trading off Ingest Cost and Query Latency:} Among the combination of values that meet the precision and recall targets, the selection is based on {\em balancing} the ingest- and query-time costs. For example, picking a model CheapCNN$_i$ that is more accurate will have higher ingest cost, but lower query cost because we can use a lower {\sf K}. Using a less accurate CheapCNN$_i$ will have the opposite effect. 
{\name} identifies ``intelligent defaults'' that sharply improve one of the two costs for a small worsening of the other cost.

Figure~\ref{fig:parameter_selection} illustrates the parameter selection based on the ingest cost and query latency for one of our video streams (\video{auburn\_c}). The figure plots all the viable ``configurations'' (i.e., set of parameters that meet the precision and recall target) based on their ingest cost (i.e., cost of CheapCNN$_i$) and query latency (i.e., the number of clusters according to $K, L_s, T$). We first draw the \emph{Pareto boundary}~\cite{pareto}, which is the set of configurations that cannot improve one metric without worsening the other. {\name} can discard all the other configurations because at least one point on the Pareto boundary is better than them in both metrics. {\name} balances between the ingest cost and query latency (\sys{Balance} in Figure~\ref{fig:parameter_selection}) by selecting the configuration that minimizes the \emph{sum of ingest and query cost} (measured in total GPU cycles).

{\name} also allows for other configurations based on the application's preferences and query rates. \sys{Opt-Ingest} minimizes the ingest cost and is applicable when the application expects most of the video streams to not get queried (such as a surveillance cameras), as this policy also minimizes the amount of wasted ingest work. On the other hand, \sys{Opt-Query} minimizes query latency even if it incurs a heavy ingest cost. Such flexibility allows {\name} to fit different applications.

%

\begin{figure}[t]
  \centering
  \includegraphics[width=0.48\textwidth]{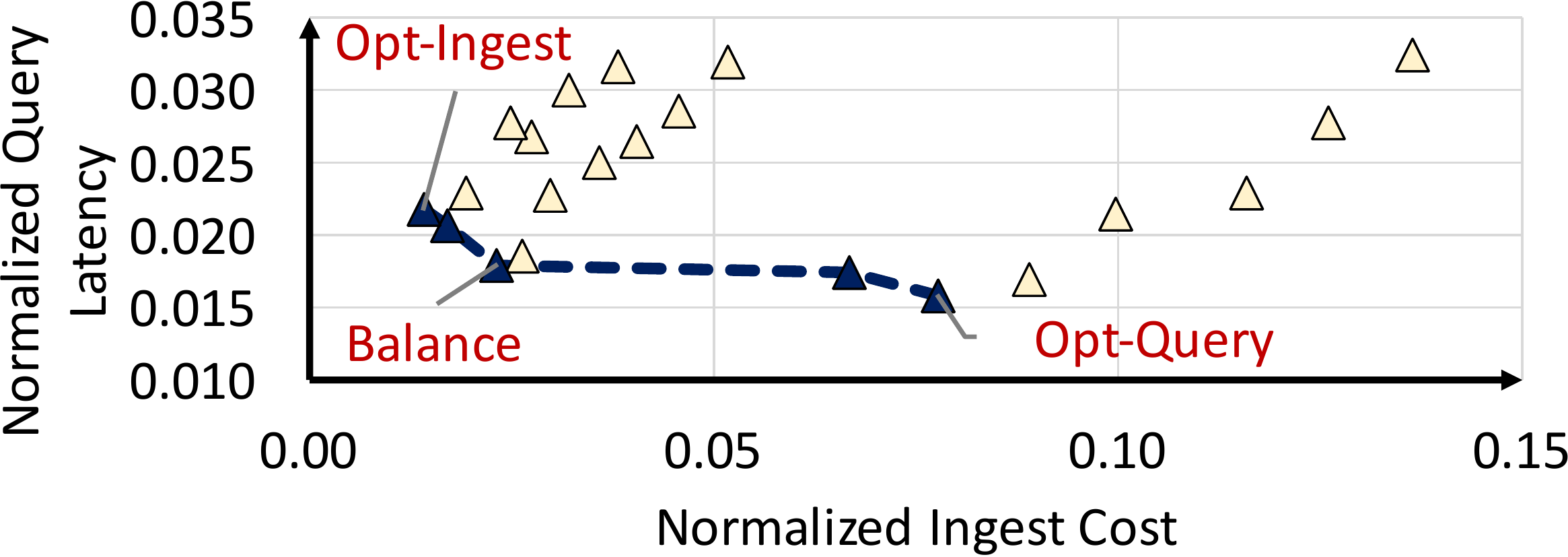}
  \caption{Parameter selection based on trading off ingest cost and query latency. The ingest cost is normalized to ingesting all objects with ResNet152, while the query latency is normalized to the time for querying all objects with ResNet152. The dashed line is the Pareto boundary.}
  \label{fig:parameter_selection}
\end{figure}

\section{Implementation Details}
\label{sec:implementation}

We describe the key aspects in {\name}'s implementation.

\noindent\textbf{Worker Processes.}
{\name}'s ingest-time work is distributed across many machines, with each machine running one {\em worker} process for {\em each video stream's} ingestion. The ingest worker receives the live video stream, 
and extracts the moving objects (using background subtraction~\cite{bgs}); it is extensible to plug in any other object detector.
The detected objects are sent to the ingest-time \cnn to infer the top-K classes and the feature vectors. The ingest worker uses the features to cluster objects in its video stream and stores the top-K index in MongoDB~\cite{mongodb} for efficient retrieval at query-time.

Worker processes also serve queries by fetching the relevant frames off the top-{\sf K} index database and classifying the objects with GT-\cnn. We parallelize a query's work across many worker processes if resources are idle.


\noindent{\bf GPUs for \cnn classification.} The cheap \cnns and GT-CNN execute on GPUs (or other hardware accelerators for \cnns) which could either be local on the same machine as the worker processes or ``disaggregated'' on a remote cluster. This detail is abstracted away from our worker process and it seamlessly works with both designs.

\noindent\textbf{Dynamically adjusting $K$ at query-time.}
As an enhanced technique, we can select a new $K_x \le K$ at query time and only extract clusters where class $X$ appears among the top-$K_x$ classes; this will result in fewer clusters and thus also lower query-time latency.
This technique is useful in two scenarios:
1) some classes might be very accurately classified by the cheap \cnn; using a lower $K_x$ will still meet the user-specified accuracy, yet will result in much lower latency;
2) if we want to retrieve only \emph{some objects of class X}, we can use very low $K_x$ to quickly retrieve them. If more objects are required, we can increase $K_x$ to extract a new batch of results.


%
%
%

\section{Evaluation}
\label{sec:evaluation}

We evaluate the {\name} prototype with more than 150 hours of videos
from 13 real video streams that span across traffic cameras,
surveillance cameras, and news channels. Our highlights are:

\begin{enumerate}
  \item On average, {\name} is simultaneously 58$\times$ (up to
    98$\times$) cheaper than the \sys{Ingest-all} baseline in its GPU
    consumption and 37$\times$ (up to 57$\times$) faster than the
    \sys{Query-all} baseline in query latency, all the while achieving
    at least 95\% precision and recall
    (\xref{sec:overall_performance}, \xref{sec:performance_by_technique}). 
    
\item {\name} provides a rich trade-off space between ingest cost and
  query latency. Among the video streams, the ingest cost is up to
  141$\times$ cheaper than the \sys{Ingest-all} baseline (and reduces
  query latency by 46$\times$) if optimizing for low-cost ingest. The
  query latency is reduced by up to 66$\times$ (with 11$\times$
  cheaper ingest) if optimizing for query latency
  (\xref{sec:ingest_query_trade_off}).
\item {\name} is effective under broad conditions such as high
  accuracy targets (\tr{one order-of-magnitude savings even for 99\%
    accuracy target}, \xref{sec:result_accuracy_target}) and various
  frame sampling rates (\tr{30 fps-1 fps}, \xref{sec:result_frame_sampling}).
\end{enumerate}

\subsection{Setup}
\label{sec:methdology}

\noindent{\bf Software Tools.} 
We use OpenCV 3.2.0~\cite{opencv} to decode the videos into frames,
and then use the built-in background subtraction
algorithm~\cite{DBLP:conf/avbs/KaewTraKulPongB15} in OpenCV to extract
moving objects from video frames. We use background subtraction
instead of object detector \cnns (e.g.,
YOLOv2~\cite{DBLP:journals/corr/RedmonF16} or Faster
R-CNN~\cite{DBLP:conf/nips/RenHGS15}) to detect objects because: (1)
running background subtraction is orders of magnitude faster than
running these \cnns, and (2) background subtraction can detect moving
objects more reliably, \tr{while object detector CNNs usually have difficulties on
small objects~\cite{DBLP:conf/eccv/LiuAESRFB16}. Nonetheless, our system can seamlessly use object detector CNNs as well.}  
We run and train \cnns with Microsoft Cognitive Toolkit
2.1~\cite{cntk}, an open-source deep learning system.


\noindent{\bf Video Datasets.} 
We evaluate 13 live video streams that span across traffic cameras,
surveillance cameras, and news channels. We evaluate each video stream
for 12 hours, which evenly cover day time and night
time. Table~\ref{table:video_dataset} summarizes the video
characteristics. By default, we evaluate each video at 30 fps and also
evaluate the sensitivity to other frame rates
(\xref{sec:result_frame_sampling}). In some figures we only show a
representative sample of 9 cameras to improve legibility.

\setlength\tabcolsep{3pt}

\begin{table}[t!]
\scriptsize \centering \caption{Video dataset characteristics}
\vspace{-10pt}
\label{table:video_dataset}
\begin{tabular}{|l|l|l|l|} \hline

Type & Name & Location & Description \\ \hline

    & auburn\_c & AL, USA & \specialcell{A commercial area intersection\\ in the City of Auburn~\cite{auburn_c}} \\ \cline{2-4}
\multirow{6}{*}{\specialcell{\\Traffic}}
    & auburn\_r & AL, USA & \specialcell{A residential area intersection\\ in the City of Auburn~\cite{auburn_r}} \\ \cline{2-4}
    & city\_a\_d & USA & \specialcell{A downtown intersection in\\
  City A\footnote{\label{citya}The video streams are obtained from real and operational
    traffic cameras in a city. We mask the city name for anonymity.}} \\ \cline{2-4}
    & city\_a\_r & USA & \specialcell{A residential area intersection\\ in City A\footnotemark[\getrefnumber{citya}]} \\ \cline{2-4}
    & bend & OR, USA & \specialcell{A road-side camera in the City\\ of Bend~\cite{bend}} \\ \cline{2-4}
    & jacksonh & WY, USA & \specialcell{A busy intersection (Town\\ Square) in Jackson Hole~\cite{jacksonh}} \\ \hline

\multirow{4}{*}{\specialcell{\\Surveillance}}
    & \specialcell{church\_st} & VT, USA & \specialcell{A video stream rotates among \\cameras in a shopping mall\\ (Church Street Marketplace)~\cite{church_st}} \\ \cline{2-4}
    & \specialcell{lausanne} & Switzerland & \specialcell{A pedestrian plazalatency (Place de \\la Palud) in Lausanne~\cite{lausanne}} \\ \cline{2-4}
    & \specialcell{oxford} & England & \specialcell{A bookshop street in the \\University of Oxford~\cite{oxford}} \\ \cline{2-4}
    & sittard & Netherlands & \specialcell{A market square in Sittard~\cite{sittard}} \\  \hline

\multirow{3}{*}{\specialcell{News}}
    & cnn & USA & \specialcell{News channel} \\ \cline{2-4}
    & foxnews & USA & \specialcell{News channel} \\ \cline{2-4}
    & msnbc & USA & \specialcell{News channel} \\ \hline

\end{tabular}
\end{table}

\noindent{\bf Accuracy Target.}  We use ResNet152, a state-of-the-art
\cnn, as our ground-truth \cnn (GT-CNN). We evaluate all extracted
objects with the GT-CNN and use the results as the correct answers.
We define a class present in a one-second segment of video if the
GT-CNN reports such class in 50\% of the frames in that segment.  We
use this criteria as our ground truth because our GT-CNN (ResNet152)
sometimes gives different answers to the exact same object in
consecutive frames, and this criteria can effectively eliminate these
random, erroneous results. We set our default accuracy target as 95\%
recall and 95\% precision. We also evaluate the results with other
accuracy targets such as 97\%, 98\% and 99\%
(\xref{sec:result_accuracy_target}). Note that in most practical
cases, only one of the two metrics (recall or accuracy) needs to be
high. For example, an investigator cares about high recall,
and looking through some irrelevant results is an acceptable trade-off.
By setting both targets high,
we are lower bounding the performance improvements that {\name} can achieve.

\noindent{\bf Baselines.}  We use two baselines for
comparisons: (1) \sys{Ingest-all}, the baseline system that uses
GT-CNN to analyze all objects at ingest time, and stores the inverted
index for query; and (2) \sys{Query-all}, the baseline system that
simply extracts objects at ingest time, and uses GT-CNN to analyze all
the objects that fall into the query interval at query time. \tr{Note
  that we strengthen both baselines with basic motion detection
  (background subtraction). Therefore, the baselines \emph{do not} run any GT-CNN on
  the frames that have no moving objects. Note that not running GT-CNN on frames with no moving objects is one of the
  core techniques in the recent NoScope 
  work~\cite{DBLP:journals/pvldb/KangEABZ17}}.

\noindent{\bf Metrics.} We use two performance metrics. The first metric is \emph{ingest
  cost}, which is the GPU time to ingest each video.  The second
metric is \emph{query latency}, which is the latency for an object
class query.  Specifically, for each video stream, we evaluate all
dominant object classes and take the average of their latencies.
(Querying for non-dominant ``OTHER'' classes is much cheaper than
querying popular classes, and would skew the results because there are
far more such classes; thus, we focus on the popular ones.)
Both metrics include only GPU time spent classifying
images and exclude other (CPU) time spent decoding video frames,
detecting moving objects, recording and loading video, and reading and
writing to the top-{\sf K} index.  We focus solely on GPU time because
when the GPU is involved, it is the bottleneck resource.
The query latency of \sys{Ingest-all} is 0 and the ingest cost of
\sys{Query-all} is 0.

\noindent{\bf Experiment Platform.} 
We run the experiments on our local cluster. Each machine in the
cluster is equipped with a state-of-the-art GPU (NVIDIA GTX Titan X),
16-core Intel Xeon CPU (E5-2698), 64 GB RAM, a 40 GbE NIC, and runs
64-bit Ubuntu 16.04 LTS.

\ignore{ 
\begin{itemize}
\item experiment setup (software / hardware)
\item dataset (video characteristics, length, frame rate)
\item CNN models (Expensive and cheap CNN models)
\item default accuracy target (10\% or 5\% FNR and FPR)
\item default ingest/query trade-off point (middle point of the Pareto
  boundary)
\end{itemize}
}

\vspace{2ex}
\subsection{End-to-End Performance}
\label{sec:overall_performance}

We first show the end-to-end performance of {\name} by showing its
ingest cost and query latency when {\name} aims to balance these two
metrics (\xref{subsec:tuning}).  Figure~\ref{fig:overall_results}
compares the ingest cost of {\name} with \sys{Ingest-all} and the
query latency of {\name} with \sys{Query-all}.
We make two main observations.

\begin{figure}[t!]
  \centering
  \includegraphics[width=0.48\textwidth]{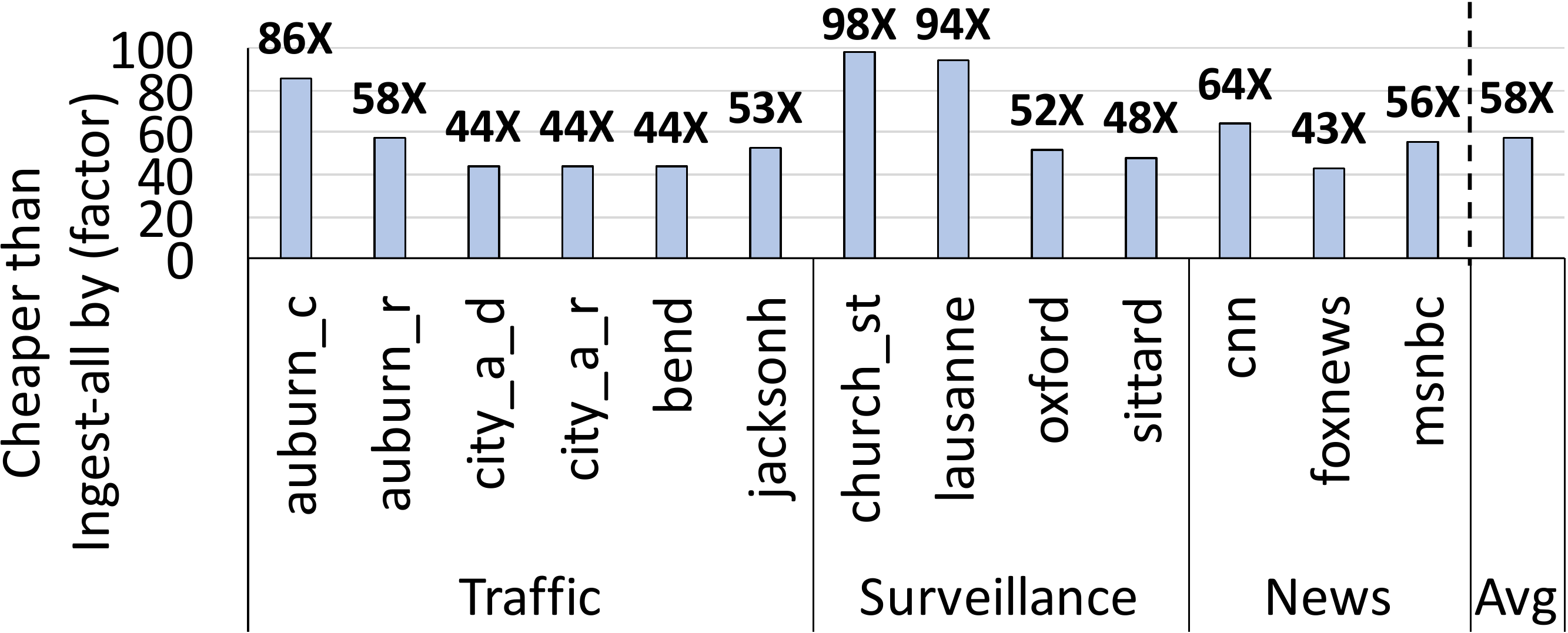}
%
\mbox{}\\
  \includegraphics[width=0.48\textwidth]{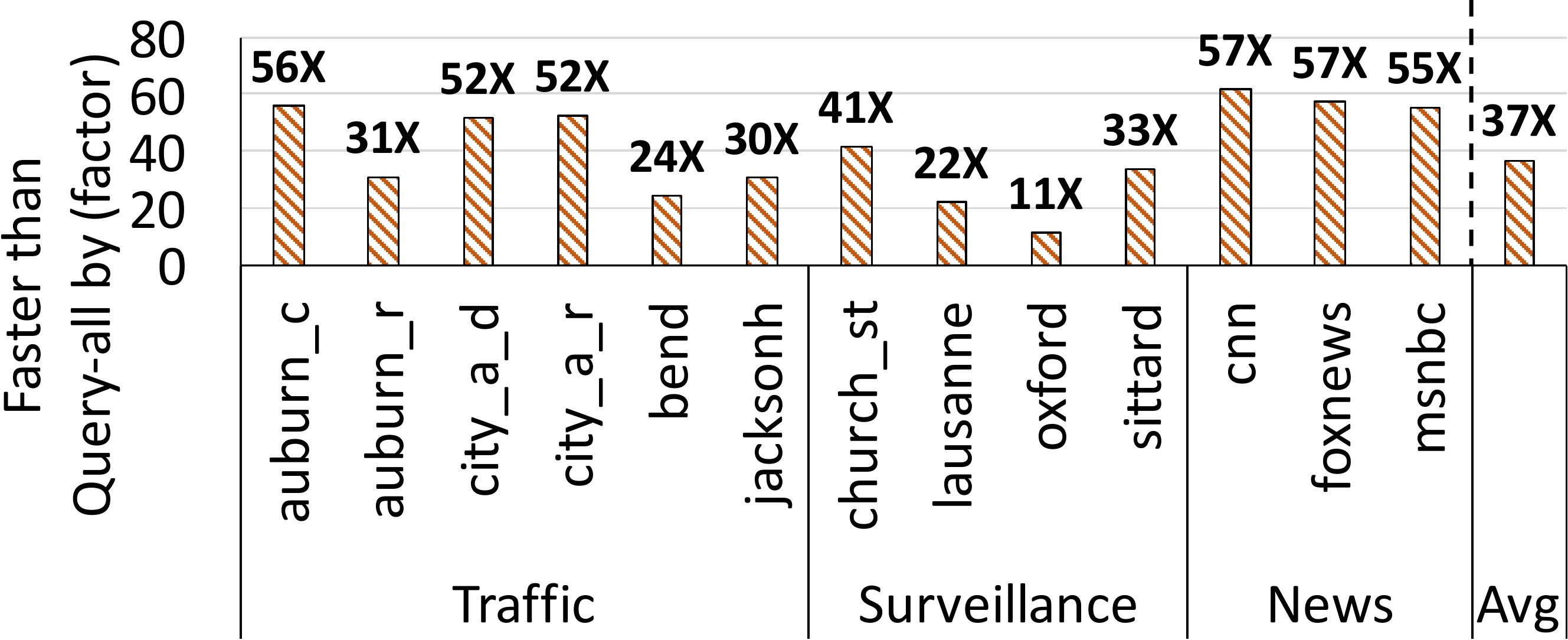}
  \caption{(Top) {\name} ingest cost compared to \sys{Ingest-all}.
    (Bottom) {\name} query latency compared to \sys{Query-all}.}
    \label{fig:overall_results}
\end{figure}

First, {\name} significantly improves query latency with a very small
ingest cost. {\name} makes queries by an average of 37$\times$ faster
than \sys{Query-all} with a very small cost at ingest time (an
average of 58$\times$ cheaper than \sys{Ingest-all}). With a 10-GPU
cluster, the query latency on a 24-hour video goes down from one hour
to less than two minutes. The processing cost of each video stream
also goes down from \$250/month to \$4/month. This shows that
{\name} can strike a very good balance between these two competing
goals very effectively.

Second, {\name} is effective across different video streams with
various characteristics. It makes queries 11$\times$ to 57$\times$
faster with a very small ingest time cost (48$\times$ to 98$\times$
cheaper) across busy intersections (\video{auburn\_c},
\video{city\_a\_d} and \video{jacksonh}), normal intersections or
roads (\video{auburn\_r} and \video{city\_a\_r}, \video{bend}),
rotating cameras (\video{church\_st}), busy plazas (\video{lausanne}
and \video{sittard}), a university street (\video{oxford}), and
different news channels (\video{cnn}, \video{foxnews}, and
\video{msnbc}). Among these videos, the gains in query latency are
smaller for relatively less busy videos (\video{auburn\_r},
\video{bend}, \video{lausanne}, and \video{oxford}). This is because
these videos are dominated by fewer object classes, and {\name} has
more work (i.e., analysis using GT-CNN) to do at query time for these
classes. We conclude that the core techniques of {\name}
are general and effective on a variety of real-world videos. 

\ignore{ 
\begin{itemize}
  \item We strike a good balance between ingest cost and query
    latency. By investing very small cost at ingest time (up to 208X
    cheaper than the ingest-all baseline), we can save query latency
    by up to 69X
  \item We show that such good balance can be achieved across various
    videos. We can get good results even when the videos have many
    different classes of objects (CNN / Fox News) or very busy
    (Jackson Hole)
\end{itemize}
}

\subsection{Effect of Different {\nameSec} Components}
\label{sec:performance_by_technique}

Figure~\ref{fig:breakdown_by_technique} shows the breakdown of
ingest-time cost and query latency across different design
points of {\name}: (1) \sys{Compressed model}, which applies a generic compressed
model for indexing at ingest time, (2) \sys{Compressed +
  Specialized model}, which uses a per-stream specialized and compressed
model for indexing, and (3) \sys{Compressed + Specialized
  model + Clustering}, which adds feature-based clustering at ingest
time to reduce redundant work at query time.  All of the above include
the top-{\sf K} index and using GT-\cnn at query-time, and achieve the same
accuracy of 95\%. Three main observations are in order.


First, generic compressed models provide benefits for both ingest
cost and query latency, but they are not the major source of
improvement. This is because the accuracy of a generic compressed model
degrades significantly when we remove convolutional layers. In
order to retain the accuracy target, we need to choose relatively
expensive compressed models (CheapCNN$_i$) and a larger {\sf K}, which incur higher
ingest cost and query latency.

Second, specializing the model (in addition to compressing it) greatly
reduces ingest cost and query latency. Because of fewer convolutional
layers and smaller input resolution, our specialized models are
7$\times$ to 71$\times$ cheaper than the GT-CNN, while retaining the
accuracy target for each video streams. Running a specialized model at ingest
time speeds up query latency by 5$\times$ to 25$\times$
(Figure~\ref{fig:query_breakdown_by_technique}).

Third, clustering is a very effective technique to further reduce
query latency with unnoticeable costs at ingest time. As
Figure~\ref{fig:query_breakdown_by_technique} shows, using clustering
(on top of a specialized compressed model) reduces the query latency
by up to 56$\times$, significantly better than just running a
specialized model at ingest time. This gain comes with a negligible
cost (Figure~\ref{fig:ingest_breakdown_by_technique}), because we run
our clustering algorithm (\xref{subsec:redundancy}) on the CPUs of the
ingest machine, which is fully pipelined with the GPUs that run the
specialized \cnn model.

\begin{figure}[t!]
 \centering
 \begin{subfigure}[t]{0.50\linewidth}
 \centering
 \includegraphics[width=1.0\textwidth]{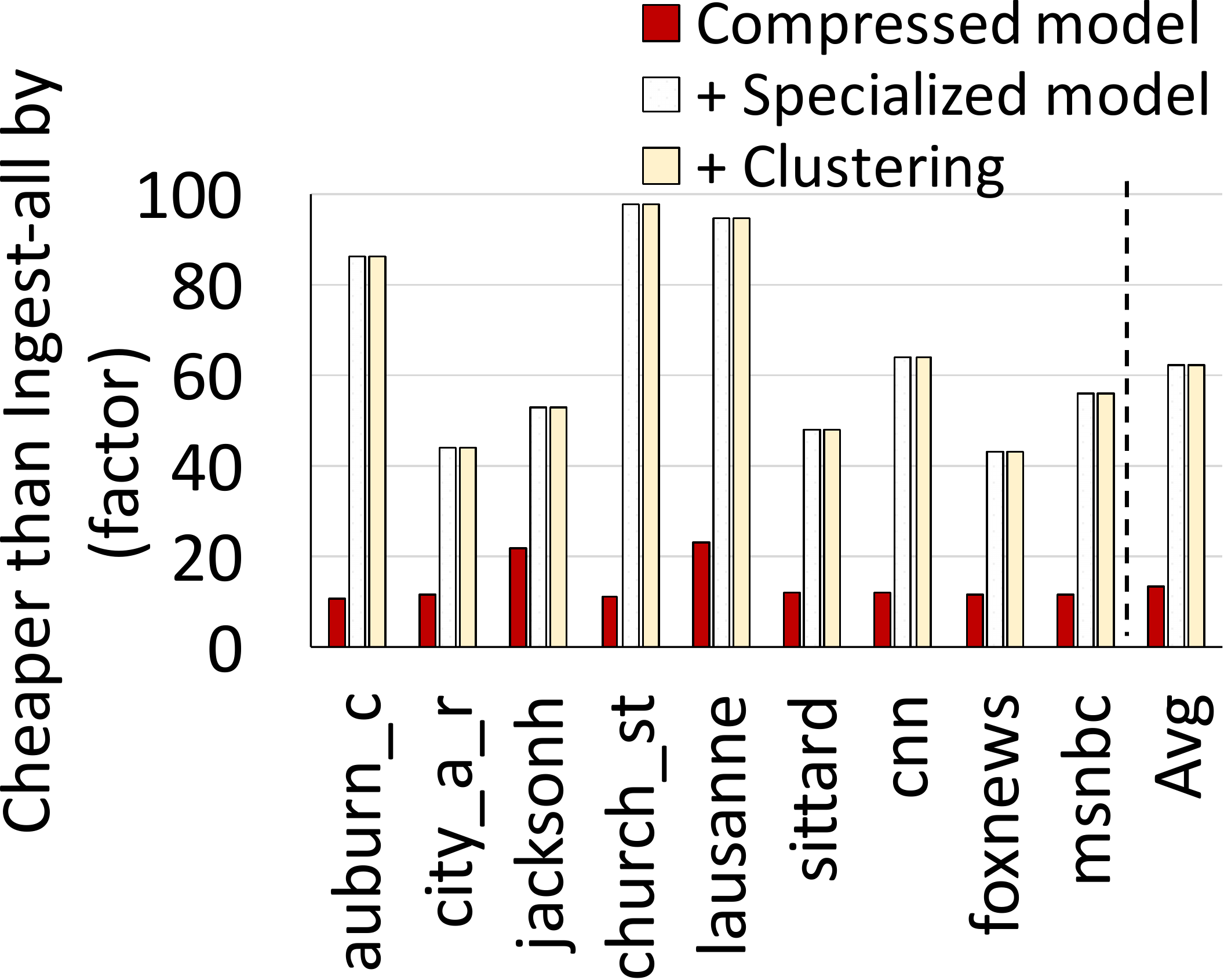}
 \caption{Ingest cost}
 \label{fig:ingest_breakdown_by_technique}
 \end{subfigure}
 \begin{subfigure}[t]{0.48\linewidth}
 \centering
 \includegraphics[width=1.0\textwidth]{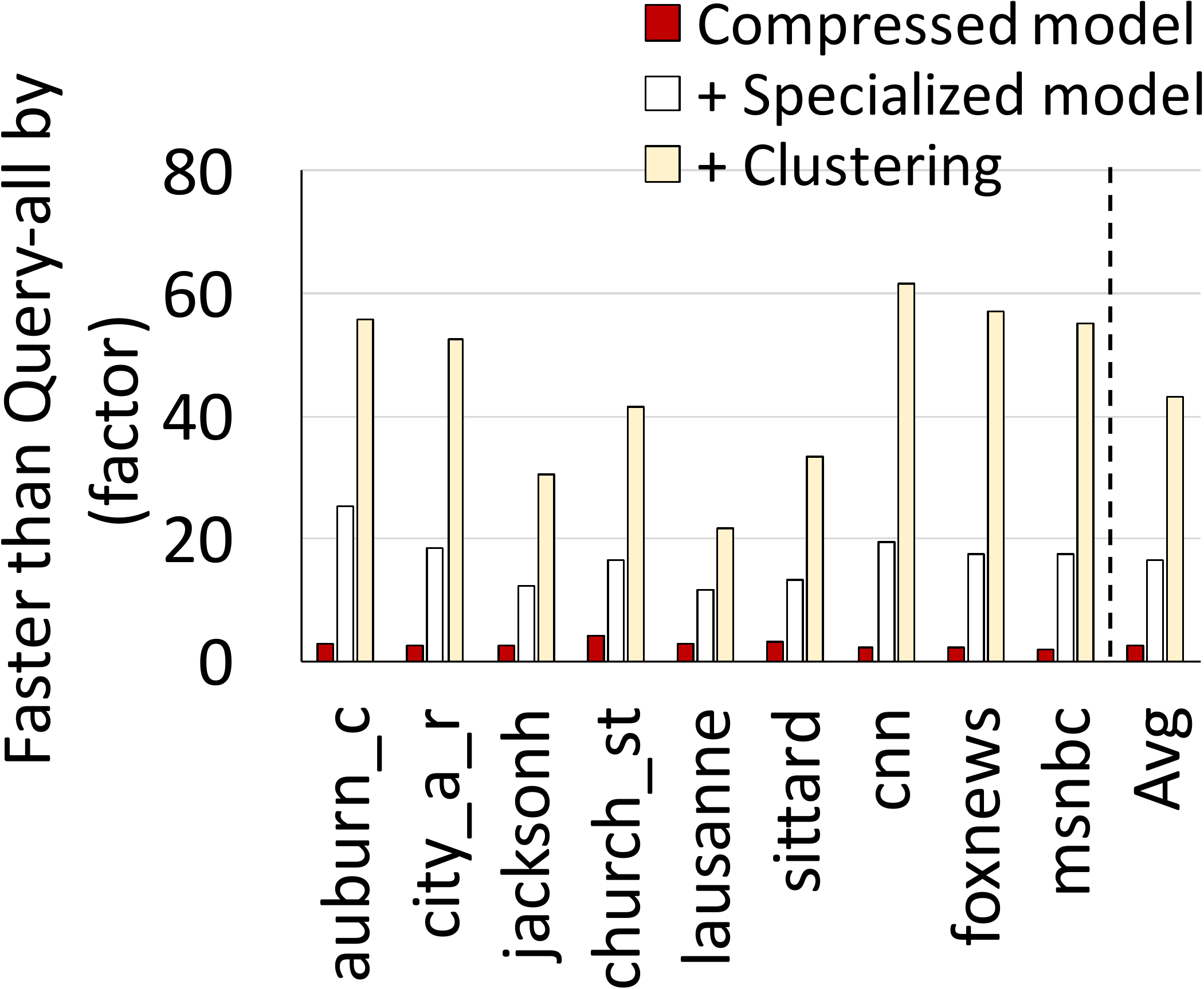}
 \caption{Query latency}
 \label{fig:query_breakdown_by_technique}
 \end{subfigure}
 \caption{Effect of different {\name} components}
 \label{fig:breakdown_by_technique}
\end{figure}

\ignore{ 
We will show two figures (ingest and query improvement) to break it
down:

For both figures, we show the results for
\begin{enumerate}
\item baseline
\item pixel-diff only
\item pixel-diff + generic cheap model + clustering
\item pixel-diff + specialized cheap model
\item pixel-diff + specialized cheap model + clustering
\end{enumerate}

Key takeaways:
\begin{itemize}
  \item Pixel-diff can help both ingest and query, but it's not the
    major improvement source. It is different from prior work because our
    baseline has background subtraction already, so frames without any
    object are not considered even for the baseline.
  \item Specialization is a very important technique. It is because
    specialized models only need to distinguish the dominant classes
    in a specific video stream, so it can achieve similar accuracy
    with fewer layers and smaller input size.
  \item Clustering is a very useful technique to remove the
    redundancy, and it is relatively cheap to do.
\end{itemize}

}

\subsection{Ingest Cost vs. Query Latency Trade-off}
\label{sec:ingest_query_trade_off}

One of the interesting features of {\name} is the flexibility to tune
its system parameters to achieve different application goals
(\xref{subsec:tuning}).  Figure~\ref{fig:result_trade_off} from
\xref{sec:intro} depicted three alternative settings for {\name} that
illustrate the trade-off space between ingest cost and query latency,
using the \video{auburn\_c} video stream:
(1) \sys{{\nameTxt}-Opt-Query}, which optimizes for query
latency by increasing ingest cost, (2) \sys{{\nameTxt}-Balance}, which
is the default option that balances these two metrics
(\xref{subsec:tuning}), and (3): \sys{{\nameTxt}-Opt-Ingest}, which is
the opposite of \sys{{\nameTxt}-Opt-Query}.  The results are'shown
relative to the two baselines.
The chart at the right of the figure is the zoomed-in
region that covers the three settings of {\name}, and each data
label $(I, Q)$ indicates its ingest cost is $I\times$ cheaper than
\sys{Ingest-all}, while its query latency is $Q\times$ faster than
\sys{Query-all}.



As Figure~\ref{fig:result_trade_off} shows, {\name} offers very good
options in the trade-off space between ingest cost and query
latency. \sys{{\nameTxt}-Opt-Ingest} achieves 141$\times$ cheaper
cost than \sys{Ingest-all} to ingest the video stream, and makes the
query 46$\times$ faster than doing nothing at ingest
(\sys{Query-all}). On the other hand, \sys{{\nameTxt}-Opt-Query}
reduces query latency by 63$\times$ with a relatively higher ingest
cost, but it is still 26$\times$ cheaper than \sys{Ingest-all}. As
they are all good options compared to the baselines, such flexibility
allows a user to tailor {\name} for different contexts. For example,
a traffic camera that requires fast turnaround time for queries can
use \sys{{\nameTxt}-Opt-Query}, while a surveillance video stream that will be queried very rarely would
choose \sys{{\nameTxt}-Opt-Ingest} to reduce the amount of wasted ingest cost.


Figure~\ref{fig:result_trade_off_all} shows the $(I, Q)$ values for
both \sys{{\nameTxt}-Opt-Ingest} (\sys{Opt-I}) and
\sys{{\nameTxt}-Opt-Query} (\sys{Opt-Q}) for the representative
videos. As the figure show, the trade-off flexibility exists among all
the other videos. On average, \sys{{\nameTxt}-Opt-Ingest} spends only
95$\times$ cheaper ingest cost to provide 35$\times$ query latency
reduction. On the other hand, \sys{{\nameTxt}-Opt-Query} makes queries
49$\times$ faster with a higher ingest cost (15$\times$ cheaper
than \sys{Ingest-all}). We conclude that {\name} provides good
flexibility between ingest cost and query latency, and makes it a
better fit in different contexts.

\begin{figure}[h]
  \centering
  \includegraphics[width=0.48\textwidth]{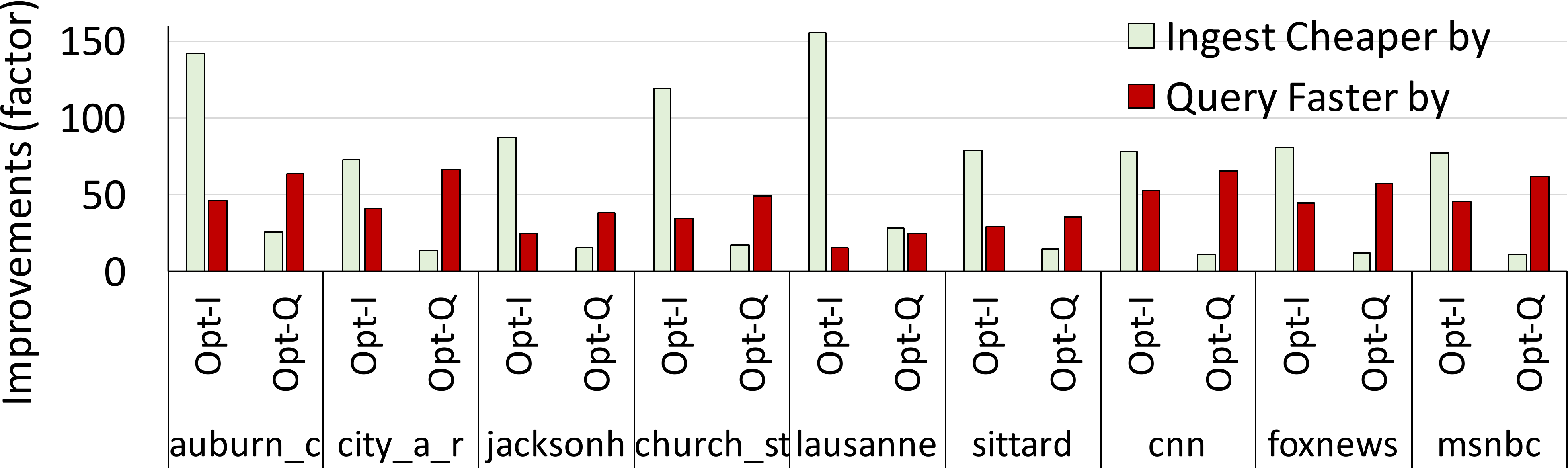}
  \caption{Trade-offs between ingest cost and query latency}
  \label{fig:result_trade_off_all}
\end{figure}
\vspace{-10pt}

%
%


\subsection{Sensitivity to Accuracy Target}
\label{sec:result_accuracy_target}

Figures~\ref{fig:result_accuracy_ingest}
and~\ref{fig:result_accuracy_query} illustrate the improvements of
ingest cost and query latency of {\name} compared to the baselines
under different accuracy targets. Other than the default 95\% accuracy
target (recall and precision), we evaluate three higher targets, 97\%,
98\%, and 99\%.

As the figures show, with higher accuracy targets, the ingest costs are
about the same, and the improvement of query latency
decreases. {\name} keeps the ingest cost similar (62$\times$ to
64$\times$ cheaper than the baseline) because it still runs the
specialized and compressed \cnn at ingest time. However, when the
accuracy targets are higher, {\name} needs to select more top-{\sf K}
classification results, which increases the work at
query time. On average, the query latency of {\name} is faster than
\sys{Query-all} by 15$\times$, 12$\times$, and 8$\times$ with respect
to 97\%, 98\%, and 99\% accuracy targets. We conclude that the
techniques of {\name} can achieve higher accuracy targets with
significant improvements on both ingest cost and query latency.

\begin{figure}[h]
  \centering
  \includegraphics[width=0.48\textwidth]{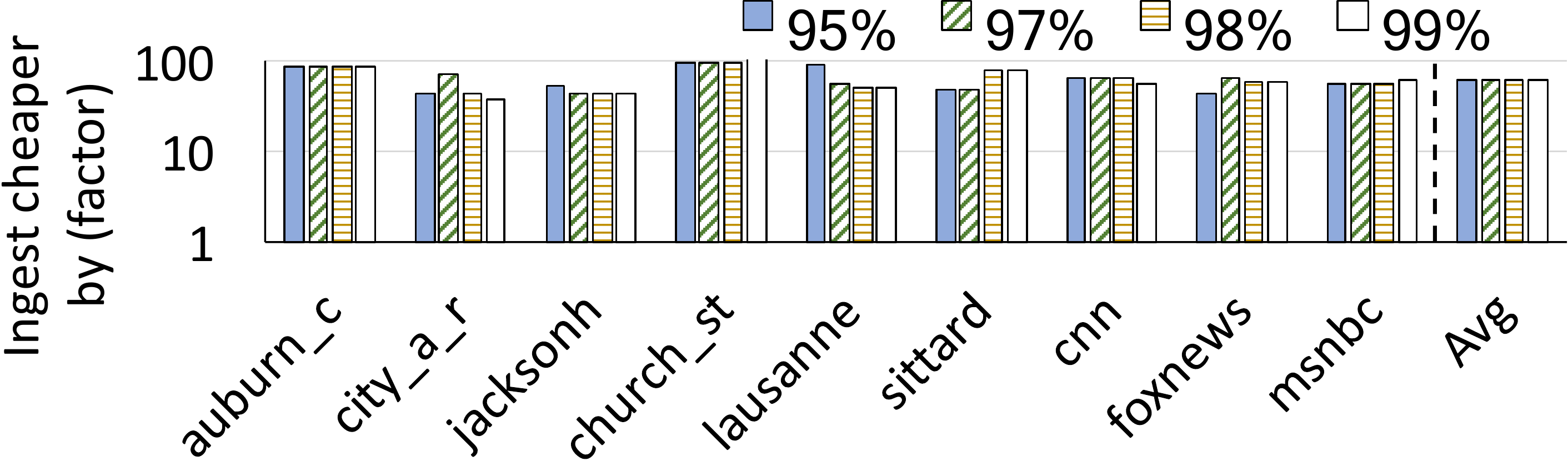}
  \caption{Ingest cost sensitivity to accuracy target}
  \label{fig:result_accuracy_ingest}
\end{figure}

\begin{figure}[h]
  \centering
  \includegraphics[width=0.48\textwidth]{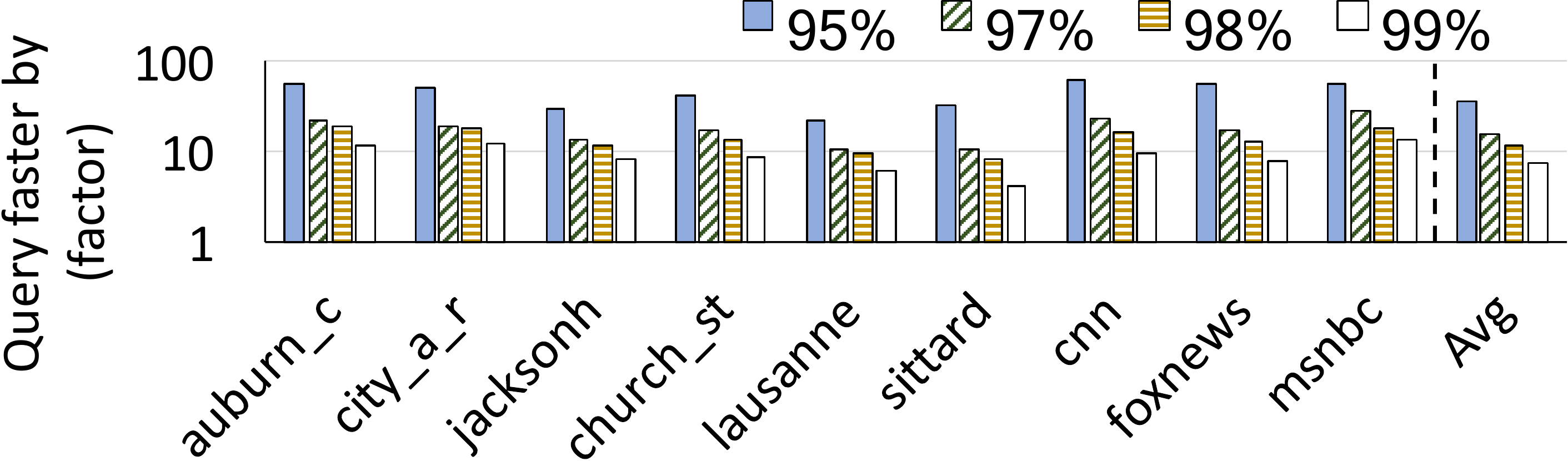}
  \caption{Query latency sensitivity to accuracy target}
  \label{fig:result_accuracy_query}
\end{figure}

\subsection{Sensitivity to Frame Sampling}
\label{sec:result_frame_sampling}

A common approach to reduce the video processing time is to use frame
sampling (i.e., periodically select a frame to process). However, not
all applications can use frame sampling because it can miss objects
that show up and disappear within a frame sampling window. As the
frame sampling rate is an application dependent choice, we study the
sensitivity of {\name}'s performance to different frame
rates. Figures~\ref{fig:result_frame_sample_ingest}
and~\ref{fig:result_frame_sample_query} show the ingest cost and query
latency of {\name} at different frame rates (i.e., 30 fps, 10 fps, 5 fps,
and 1 fps) compared to \sys{Ingest-all} and \sys{Query-all},
respectively. We make two observations.

\begin{figure}[h]
  \centering
  \includegraphics[width=0.48\textwidth]{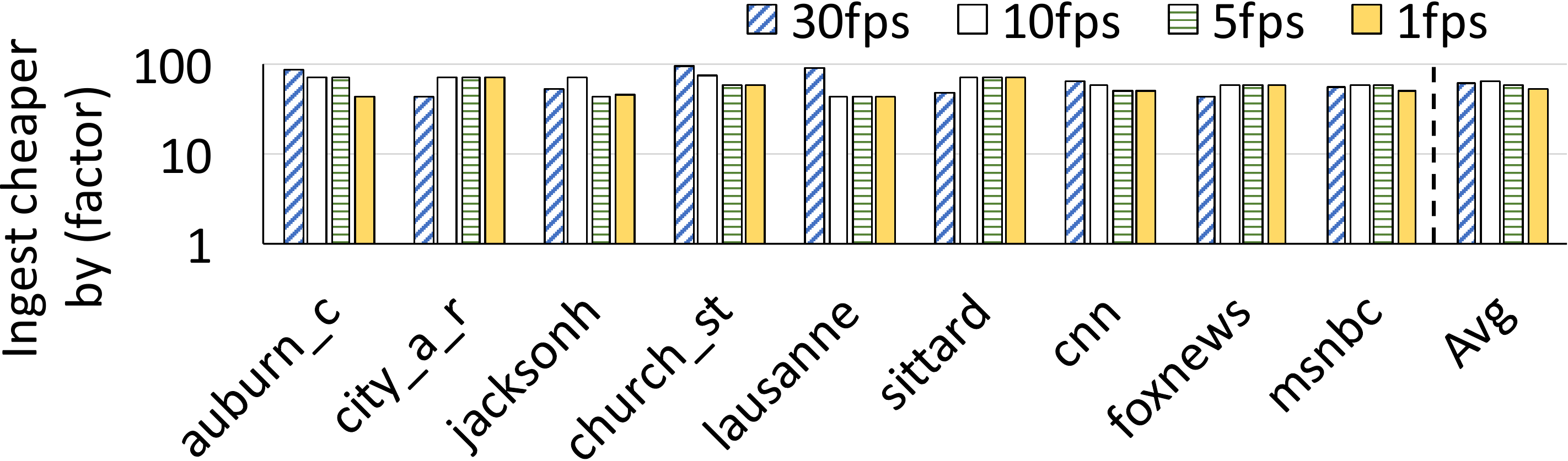}
  \caption{Ingest cost sensitivity to frame sampling}
  \label{fig:result_frame_sample_ingest}
\end{figure}


\begin{figure}[h]
  \centering
  \includegraphics[width=0.48\textwidth]{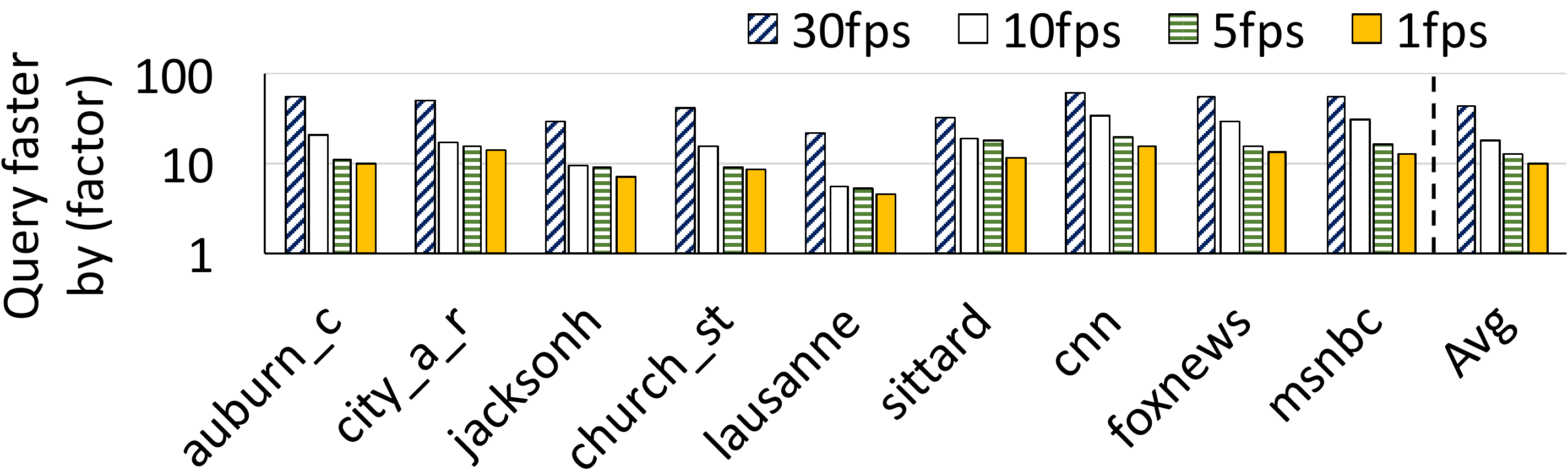}
  \caption{Query latency sensitivity to frame sampling}
  \label{fig:result_frame_sample_query}
\end{figure}

First, the ingest cost reduction is roughly the same across the different
frame rates. On average, the ingest cost of {\name} is 62$\times$
cheaper than \sys{Ingest-all} at 30 fps, and it is 64$\times$ to
58$\times$ cheaper at lower frame rates. This is because the major
ingest cost saving comes from the specialized and compressed \cnn
models (\xref{sec:performance_by_technique}), which are orthogonal to
frame sampling rates.

Second, the query latency improvement of {\name} degrades with lower
frame rates. This is expected because one of our key techniques to
reduce query latency is redundancy elimination, especially clustering
similar objects using \cnn feature vectors. At lower frame rates, the
benefit of this technique reduces because there are fewer
redundancies. Nonetheless, on average, {\name} is still one order of
magnitude faster than \sys{Query-all} at a very low frame rate (1
fps).


\ignore{
  \subsection{Effects of Ingest-time Slack}

Show the results without ingest-time optimization (everything happen
at query time). \kh{Do we need to do this? The downside is we may
  invite the head-to-head comparison with the Stanford paper}
}

\subsection{Applicability with Different Query Rate}
\label{sec:result_query_rate}

There are two factors that can affect the applicability of {\name}: 1)
the number of classes that get queried over time and 2) the fraction
of videos that get queried. In the first extreme case where all the
classes and all the videos are queried, \sys{Ingest-all} could be a
good option because its cost is amortized among all the queries. In
our study, even in such an extreme case, the overall cost of {\name}
is still 4$\times$ cheaper than \sys{Ingest-all} on average (up to
6$\times$ cheaper) because we run a very cheap \cnn at ingest time,
and we run GT-\cnn \emph{per object cluster} only \emph{once}
(\xref{sec:implementation}), so the overall cost is still cheaper than
\sys{Ingest-all}.

The second extreme case is only a tiny fraction of videos gets
queried. While {\name} can save the ingest cost by up to 141$\times$
(\xref{sec:ingest_query_trade_off}), it can be more costly than
\sys{Query-all} if the fraction of videos gets queried is less than $
\frac{1}{141} = 0.7\%$. In such a case, we can choose to do nothing at
ingest time and run all the techniques of {\name} only at query time
when we know the fraction of videos that get queried. While this
approach increases query latency, it still reduces the query
latency by a average of 22$\times$ (up to 34$\times$) than
\sys{Query-all} in our evaluation. We conclude that {\name} is
still better than both baselines even under extreme query rates.

\ignore{Show two graphs here. One is what if the number of query increases and
cover more classes. Are we still better than the ingest-all baseline?
(Because it can amortize the cost). The other one is what if the query
only happens to a very small fraction of videos. At which point our
technique would be worse than the query-all baseline? }

\ignore{
\subsection{Discussions (not part of the outline)}
when computing average improvement, do we assume that each class is
equally likely to show up in a query? what if there's a different
distribution? what gains do we get then?

\kh{yes, we assume each class is equal. we can show the variation
  among all classes}

\subsection{Misc}
\begin{itemize}
\item show that we can properly deal with the ``other'' class
\item show distribution ``across'' classes, not just average. is it
  possible that some classes do much worse than others? how to fix
  this? \kh{It's possible, but I only observe this in 5-min videos. In
  1-hour videos, it seems almost all classes have similar improvements}
\item show that by tuning the various knobs we can achieve other tradeoffs in the design space
\item design space relative to ingest cost and query rate
\item show that feature clustering is much better than pixel-level clustering/dedup
\end{itemize}
}

\section{Related Work}
\label{sec:related}

\tr{To our best knowledge, {\name} is the first system that offers low-cost, low-latency,and high-accuracy video queries by balancing between ingest-time cost and query latency.}
We now discuss
work related to our key techniques.

\noindent{\bf 1) Cascaded classification.}
Various works in vision research propose speeding up classification by cascading a
series of classifiers. Viola et al.~\cite{DBLP:conf/cvpr/ViolaJ01} is the earliest work which
cascades a series of classifiers (from the simplest to the most
complicated) to quickly disregard regions in an image. Many improvements followed (e.g., ~\cite{DBLP:conf/icip/LienhartM02,
  DBLP:journals/tkde/YangLCP06, DBLP:conf/icml/XuKWC13}). \cnns are also cascaded (e.g.,
~\cite{DBLP:conf/cvpr/SunWT13, DBLP:conf/cvpr/LiLSBH15, mcdnn,
  DBLP:conf/iccv/CaiSV15}) to reduce object
detection latency. 
Our work is different in two major ways. First, we \emph{decouple} the compressed
\cnn from the GT-CNN, which allows us to choose from a wider range for
ingest-time \cnns and allows for better trade-offs
between ingest cost and query latency, a key aspect of our
work. Second, we cluster similar objects using \cnn features to
eliminate redundant work, which is a new and effective
technique for video streams.

\noindent{\bf 2) Neural network compression.}
Recent work proposes various
techniques to reduce the running time of \cnns. These techniques include shallow
models~\cite{DBLP:conf/nips/BaC14}, predicting
weights~\cite{DBLP:conf/nips/DenilSDRF13}, matrix
pruning~\cite{pruning1,pruning2}, model quantization~\cite{DBLP:conf/iclr/HanMD16}, and others
(e.g.,~\cite{lowrank, fitnets, DBLP:conf/nips/DentonZBLF14,
  DBLP:conf/sips/HwangS14, DBLP:conf/icassp/AnwarHS15,
  DBLP:journals/corr/HintonVD15, DBLP:conf/eccv/RastegariORF16}). 
Our work is largely orthogonal to these, in that our system
is not tied to a specific model compression technique, and we can
employ any of these techniques.

\noindent{\bf 3) Context-specific model specialization.}
Context-specific specialization of models can improve
accuracy~\cite{DBLP:conf/dicta/MhallaMCGA16} or reduce running
time~\cite{mcdnn, DBLP:journals/pvldb/KangEABZ17,
  DBLP:journals/corr/ShenHPK17}. Among these, the closest to our work
is Kang et al.'s proposal,
NoScope~\cite{DBLP:journals/pvldb/KangEABZ17}, which aims to optimize \cnn-based video queries. A few key differences stand out. 
First, NoScope applies all the optimizations at query-time, while {\name} adopts a different architecture by splitting work between ingest- and query-time. Thus, {\name} trades off higher ingest cost for {\em even lower} query latency.
Second, NoScope optimizes \cnns for a single class, while we optimize ingest \cnns for all frequent classes in the stream and allow queries even for the rare -- OTHER -- classes. 
Finally, we use the object feature vectors to cluster similar objects and create an index to map classes to clusters; this allows us to efficiently query across {\em all} classes, while NoScope has to redo all query-time work, \tr{including training specialized CNNs}, for each query.



\noindent{\bf 4) Stream processing systems.}
Systems for general stream data processing (e.g.,~\cite{DBLP:conf/cidr/AbadiABCCHLMRRTXZ05, amini2006spc,
  DBLP:conf/sigmod/ChandrasekaranCDFHHKMRS03,
  DBLP:conf/vldb/CarneyCCCLSSTZ02, storm, DBLP:conf/vldb/TatbulCZ07,
  DBLP:conf/vldb/TuLPY06, DBLP:conf/sosp/ZahariaDLHSS13,
  DBLP:conf/nsdi/RabkinASPF14, DBLP:conf/nsdi/LinFQXYZZ16,
  DBLP:conf/sigmod/BailisGMNRS17}) 
and specific to video analytics
(e.g.,~\cite{DBLP:conf/nsdi/ZhangABPBF17}) 
mainly
focus on the general stream processing challenges such as load
shedding, fault tolerance, distributed execution, or limited network
bandwidth. In contrast, our work is specific for querying on recorded video
data with ingest and query trade-offs, thus it is mostly orthogonal
to these. We can integrate {\name} with one of these general stream
processing system to build a more fault tolerable system.

\tr{\noindent{\bf 5) Video indexing and retrieval.} A large body of
  works in multimedia and information retrieval research propose
  various content-based video indexing and retrieval techniques to
  facilitate queries on videos (e.g.,
  ~\cite{DBLP:journals/mta/SnoekW05, DBLP:journals/tomccap/LewSDJ06,
    DBLP:journals/ftir/SnoekW09, DBLP:journals/tsmc/HuXLZM11}). Among
  them, most works focus on indexing videos for different types of
  queries such as shot boundary
  detection~\cite{DBLP:journals/tcsv/YuanWXZLLZ07}, semantic video
  search~\cite{DBLP:conf/icassp/ChangMS07}, video
  classification~\cite{DBLP:journals/tsmc/BrezealeC08}, or
  spatio-temporal information-based video
  retrieval~\cite{DBLP:journals/pr/RenSSZ09}. Some works
(e.g.,~\cite{DBLP:conf/civr/ChristelC06,
  DBLP:conf/trecvid/SnoekSRHGORGWKS10}) focus on the query interface to enable query by keywords, concepts, or examples. These works are largely orthogonal to our work because we focus on the \emph{cost and latency} of video queries, not query types or interfaces. We believe our idea of splitting ingest-time and query-time work is generic for videos queries, and can be extended to different types of queries.} 

\section{Conclusion}
\label{sec:conclusion}

Answering queries of the form, {\em find me frames that contain objects of class X} is  an important workload on recorded video datasets. Such queries are used by analysts and investigators, and it is crucial to answer them with low latency and low cost. We present {\name}, a system that performs low cost ingest-time analytics on live video that later facilitates low-latency queries on the recorded videos. {\name} uses compressed and specialized \cnns at ingest-time that substantially reduces cost. It also clusters similar objects to reduce the work done at query-time, and hence the latency. {\name} selects the ingest-time \cnn and its parameters to smartly trade-off between the ingest-time cost and query-time latency. Our evaluations using $150$ hours of video from traffic, surveillance, and news domains show that {\name} reduces GPU consumption by $58\times$ and makes queries $37\times$ faster compared to current baselines. We conclude that {\name} is a promising approach to querying large video datasets. We hope that {\name} will enable future works on better determining the ingest-time and query-time trade-offs in video querying systems.
Our next steps include training a specialized and highly accurate \tr{query-time \cnn} for each stream and object to further reduce query latency.



\bibliographystyle{IEEEtranS}

\begingroup
{\small \bibliography{ref}}  
\endgroup



\end{document}